\documentclass[manuscript]{acmart}
\AtBeginDocument{%
  \providecommand\BibTeX{{%
    \normalfont B\kern-0.5em{\scshape i\kern-0.25em b}\kern-0.8em\TeX}}}

\copyrightyear{2023} 
\acmYear{2023} 
\setcopyright{acmlicensed}
\acmConference[IUI '23]{28th International Conference on Intelligent User Interfaces}{March 27--31, 2023}{Sydney, NSW, Australia}
\acmBooktitle{28th International Conference on Intelligent User Interfaces (IUI '23), March 27--31, 2023, Sydney, NSW, Australia}
\acmPrice{15.00}
\acmDOI{10.1145/3581641.3584099}
\acmISBN{979-8-4007-0106-1/23/03}

\usepackage[colorinlistoftodos,prependcaption,textsize=tiny]{todonotes}

\newcommand{\final}[1]{{\leavevmode\color{black}#1}}
\newcommand{\change}[1]{{\leavevmode\color{black}#1}}

\usepackage{soul}
\usepackage{setspace}

\begin{document}

\title[\textit{Preprint version-} SeeChart: Enabling Accessible Visualizations]{SeeChart: Enabling Accessible Visualizations Through Interactive Natural Language Interface For People with Visual Impairments}


\author{Md Zubair Ibne Alam}
\affiliation{%
  \institution{York University}
  \city{Toronto}
  \country{Canada}}
\email{zalam48@yorku.ca}

\author{Shehnaz Islam}
\affiliation{%
  \institution{York University}
  \city{Toronto}
  \country{Canada}}
\email{shen16@my.yorku.ca}

\author{Enamul Hoque}
\affiliation{%
  \institution{York University}
  \city{Toronto}
  \country{Canada}}
\email{enamulh@yorku.ca}


\newcommand{\seechart}{\textit{SeeChart}}

\begin{abstract}
  Web-based data visualizations have become very popular for exploring data and communicating insights. Newspapers, journals, and reports regularly publish visualizations to tell compelling stories with data. Unfortunately, most visualizations are inaccessible to readers with visual impairments. For many charts on the web, there are no accompanying alternative (alt) texts, and even if such texts exist they do not adequately describe important insights from charts. To address the problem, we first interviewed 15 blind users to understand their challenges and requirements for reading data visualizations. Based on the insights from these interviews, we developed \seechart, an interactive tool that automatically deconstructs charts from web pages and then converts them to accessible visualizations for blind people by enabling them to hear the chart summary as well as to interact through data points using the keyboard. Our evaluation with 14 blind participants suggests the  efficacy of \seechart\ in understanding key insights from charts and fulfilling their information needs while reducing their required time and cognitive burden.
\end{abstract}





\begin{teaserfigure}
  \centering
  \includegraphics[width=1\textwidth]{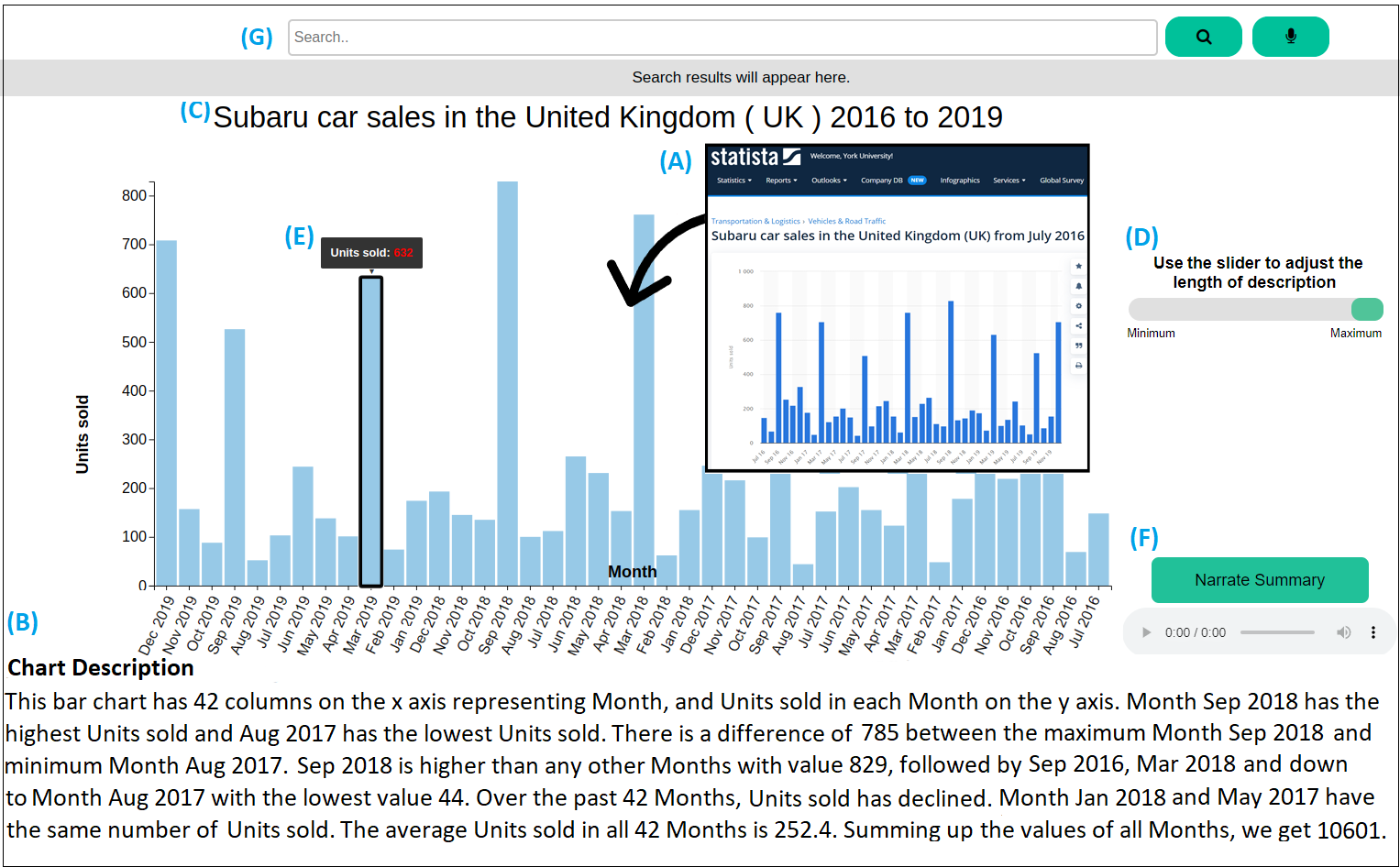}
  \caption{\seechart\ automatically deconstructs a D3 chart from a web page (A) and converts it into an accessible visualization for blind people. The tool generates and presents a textual summary from the extracted data (B) along with other metadata such as the chart title (C). Users can adjust the length of the summary (D). The interface allows users to navigate through the data points (E) via keyboard shortcuts and get audio-feedback (F). \change{The voice search feature (G) allows user to search for key statistics as well as data values and attributes
  from the chart.
  } 
  }
  \label{fig:teaser}
\end{teaserfigure}


\maketitle

\section{Introduction}

Web-based visualizations have become very popular as people regularly use them to explore datasets and communicate important insights. Visualizations frequently appear in news articles, financial reports, scientific articles, and blogs to help readers understand data. Unfortunately, people who have visual impairments often cannot interpret and understand visualizations, as there is little or no accessibility support for them~\cite{jung2021communicating, Kim2021}. Globally, there are at least 253 million people with visual impairments including 36 million people who are blind\cite{ackland2017world}.
Unlike sighted people who can explore and comprehend web-based interactive visualizations, people with visual disabilities
are often given no choice to understand charts, aggravating inequalities in access to information among individuals with disabilities in comparison to those who are not living with a disability~\cite{jung2021communicating}.

While there have been some research and development on assistive technologies for people with visual impairments, 
\change{most of them do not specifically allow blind users to interactively access the underlying data and get insights from charts}~\cite{jung2021communicating, Kim2021, accessibile-viz}. Accessible tools such as screen-readers can help people by reading alternative (alt) texts which are embedded with graphical elements in a web page. However, a survey \cite{disability2004web} found that more than 80\% of websites are not developed following standards (e.g., W3C standards \cite{W3C}) and guidelines (e.g., Web Content Accessibility Guidelines \cite{WCAG}) and do not always provide the alt text. Even when there are alt texts embedded with visualizations, they do not follow the prescribed guidelines, namely, the WCAG guidelines \cite{WCAG}, the Diagram Center’s guidelines \cite{Diagram}, the Penn State’s accessibility guidelines \cite{PENN}, and CFPB’s guidelines \cite{CFPB}. Moreover, most alt texts do not provide important information like chart type, axes, or data trends\cite{jung2021communicating} which are critical to people with blindness~\cite{accessibile-viz}. Thus, without alt text or alt text with insufficient information, people with visual impairments are unable to fulfill their information needs from charts. \change{Prior study\cite{understandingScreenReader} showed that screen-reader users can extract 61.48\% less accurate information from web-based visualization after spending 210.96\% more time than sighted people.} Beyond alt text, software tools like SAS \cite{SAS} and research prototypes (e.g., ~\cite{choi-et-al-2010, zhao2008data}) used data sonification with keyboard navigation where values of each bar are represented by the pitch of the corresponding piano note. However, studies showed that participants prefer speech over
sonification because of cognitive overload~\cite{sakhardande2019comparing} and integrating such techniques in web environment can be challenging~\cite{accessibile-viz}.

To address the aforementioned issue, we present a design study where our goal is to enable accessible visualizations by developing a web-based interactive system. We started by conducting a needfinding interview with 15 people with blindness to understand how they currently attempt to explore commonly used charts on Web and what challenges they face during such exploration. Our participants were quite familiar with screen-reader tools (e.g., JAWS \cite{JAWS}, NVDA \cite{NVDA}) as they use them to get audio versions of alt texts. The interviewees reported that chart-captions given as alt texts provide very little insight about the chart. They also confirmed from their past experience that, in most cases, alt texts for charts  do not provide any chart-specific information or insights about the data. Similarly, blind users can rarely access the data points in a chart but even if they get access, then it is difficult to get interesting patterns, trends, or outliers by navigating through data points. From the analysis of the interview transcripts, we characterize critical user requirements including 1) the need for providing a textual summary of the chart including important patterns, trends, and outliers, 2) facilitating navigation through the chart data points, and 3) enabling interactive selection to summarize a subset of the dataset.

Based on these user requirements, we distilled several design principles which guided the design of \seechart\, an interactive web-based system that facilitates users with blindness in understanding charts through automatically generated natural language descriptions and 
exploring 
the chart through the keyboard. \final{Figure \ref{fig:teaser} shows the interface of \seechart}. \seechart\ automatically detects and deconstructs a D3 chart  (e.g., bar chart, line chart, scatterplot) by extracting data-encoding marks (e.g., \textit{bar}) and encodings that describe mappings between data and visual attributes of marks (e.g., heights of \textit{bar}) using a Chrome extension \cite{Harper2014, Harper2018}. Using the deconstructed chart, the system generates an accessible version of the corresponding chart with a comprehensive audio description generated by our natural language generation method. Through audio, \seechart\ explains how the chart encodes the data as well as the important patterns, trends, and outliers in the chart to the user. Moreover, users can navigate through the chart and select a portion of the chart with keyboard shortcuts and listen to an audio description of the selected portion. We validated our system through a user study with 14 participants with blindness who utilized our system to read charts and provided us with feedback through semi-structured interviews. The quantitative analysis and qualitative feedback revealed that readers found the \seechart\ tool improved their ability to understand charts through audio and keyboard navigation, which would be very difficult otherwise with existing tools.

The primary contributions of our work are: 1) identification of key requirements of people with visual impairments for understanding charts through interviews; 2) design and development of \seechart\ which combines natural language generation algorithm with interactive chart navigation to facilitate users with blindness in exploration and analysis; and 3) a user study which provides insights into the efficacy of our tool. As secondary contributions, we discussed the lessons learned from our study including the challenges faced by users as well as possible future directions for further developing interactive natural language interfaces for supporting readers with visual disabilities.

\section{Related Work}
Our work builds on the following main areas of prior work: (1) accessible visualization, (2) deconstructing visualizations, and (3) chart summarization. 

\subsection{Accessible Visualization}

While there have been significant research and development on improving accessibility on Web, very few works have focused specifically on chart accessibility. Recently, Kim et al. \cite{Kim2021} defined the design space for chart accessibility and identified the current knowledge gap in this research space. \change{Elavsky et al. \cite{HowAccessible} presented a set of heuristics
that facilitate designers to evaluate the accessibility of interfaces developed to present visualizations.} Screen-reader tools such as JAWS \cite{JAWS}, NVDA \cite{NVDA} support people with visual impairments to navigate through web pages and read texts through keyboard shortcuts. To further improve web accessibility various guidelines have been developed. For example, W3C standards \cite{W3C} established the  Web Content Accessibility Guidelines \cite{WCAG} so that people with visual impairments can navigate through websites. The guidelines also provide examples of alternative texts which are invisible text blocks that can be embedded with image elements in HTML with the \texttt{alt} attribute. There are also specific guidelines on formulating alt text for various graphics including charts and graphs, such as the guidelines from  WCAG \cite{WCAG}, Diagram Center’s guidelines \cite{Diagram}, the Penn State’s accessibility guidelines \cite{PENN}, and CFPB’s guidelines \cite{CFPB}. Yet, many charts on the web do not accompany alt texts\change{. As a result, most of the charts are entirely inaccessible to the screen-reader tools \cite{RichScreenReaderExperiences} and even if some charts come with alt texts}, such texts often do not follow the above guidelines and fail to provide important information like chart type, axes, or data trends~\cite{jung2021communicating}. \change{Chundury et al.~\cite{UnderstandingSensory} attempted to understand how individuals with blindness perceive surroundings using non-visual senses and based on their findings they outlined several design considerations for accessible visualizations. Murphy et al. \cite{empiricalInvestigation} ran an empirical study involving 30 blind and partially sighted computer users to find out the day-to-day challenges they face while using the Internet.} 
The study results revealed that lack of \texttt{alt} tags on images and poorly labeled page items cause confusion and make it difficult to form an overview of a web page. 

A key question here is what information should be presented in a  textual summary of a chart for effective comprehension by users with blindness. A recent interview study involving people with  visual impairments revealed how they want to utilize visualizations and how they draw the image in their head by listening to alt texts \cite{jung2021communicating}. From the analysis of interview data, the authors recommended that the alt text should include chart type, axis labels, data trends, and users should have access to data points. Another study that collected natural language summaries and analyzed them with both sighted and blind users concluded that the summaries should focus more on chart's overall trends and key statistics rather than explaining its visual constructions only ~\cite{accessibile-viz}. In this work, we have considered these recommendations along with findings from our needfinding studies to inform our chart summarization approach.

There have also been some automatic techniques for navigating charts and tables through textual summaries. The iGRAPH-Lite system ~\cite{Ferres-accessibility-2013} provides chart accessibility for blind users via generating captions and supporting navigation through the chart using the keyboard. It uses templates to provide a short description of what the chart looks like. They found that the metadata such as chart title, axis labels, and chart types are important information to blind readers. EvoGraph~\cite{Sharif2018} is a jQuery plugin that aims to create screen-reader friendly charts with information like chart title, data values, and statistics such as mean and standard deviation. \change{The SIGHT system\cite{Demir2010, accessToMultiModalArticles} is an interactive accessibility system that presents brief summaries of charts with their salient features and allows users to access statistical information (e.g., extrema, change in trends, etc.) of the charts through a menu interface.}  Overall, these work focused on helping people navigate through data points through audio descriptions without providing high-level insights conveyed by the chart. \change{VoxLens \cite{voxLens} is an interactive JavaScript plug-in for developers that a descriptive summary of a chart and then allows users to interact with the visualization by using voice-activated commands. However, such tool assumes that developer will utilize the VoXlens library to create online charts. In contrast, our \seechart\ tool is a browser extension tool that automatically deconstructs SVG visualizations from a webpage to make them accessible without relying on developers to write code. Also, unlike the above body work, \seechart\ enables interactive selection and filtering using the keyboard.} 



Beyond natural language based audio feedback, others have considered sonification (the mapping of data to non-speech sounds)~\cite{voxLens, choi-et-al-2010, zhao2008data, SAS} and tactile systems like braille displays ~\cite{lundgard2019sociotechnical}. For example, SAS \cite{SAS} uses sonification with keyboard navigation where the values of each bar are represented by the pitch of the corresponding piano note. \change{Sio et al. \cite{SupportingAccessibleThroughAudioNarratives} proposed a heuristics-based approach combining sonification with textual description to generate audio data narratives from a time-series dataset and Infosonics \cite{Infosonics} presented an unconventional method to present the audio analogue of infographics combining interactive sonifications, spoken introduction, and harmonious notes. Wang et al. \cite{SeeingThroughSounds} conducted experiments to find how different auditory channels affect the perception of visualization among people with visual impairments and they reported that pitch was the most instinctive channel to understand all data types (e.g., quantitative, ordinal, and nominal).} However, data sonification can lead to cognitive overload~\cite{sakhardande2019comparing}, and integrating such techniques in online environment is challenging~\cite{jung2021communicating}. Therefore, in this work, we are focusing on natural language summaries of charts instead of sonification.

\change{Besides information visualizations, others have focused on making data tables accessible. Brown et al. \cite{brown2003design} outlined a set of design guidelines to present tables using non-speech audio. Others have utilized the automatically extracted data from chart images to navigate through the recovered data table \cite{Choi2019VisualizingFT}. While such tabular representation of a chart provides direct access to data but fails to provide a high-level overview of the chart \cite{ramloll2001using}, \cite{understandingScreenReader}. Also, when the data table of a chart is large, navigating through the raw data values through audio feedback can lead to cognitive overload~\cite{kildal2006providing, stockman2005interactive}. }

\subsection{Deconstructing Visualizations}
The goal of deconstructing a chart is to extract the underlying data, the marks (e.g., points, bars) along with their visual attributes (e.g., color, position), and visual encodings that describe the mappings between them. Early work introduced semi-automatic systems to extract the data from the bitmap chart images \cite{Revision, jung2017chartsense}. Subsequently, others proposed fully automatic chart data extraction pipeline, however, their methods rely on various heuristics which may not work for many real-world charts and the performance was limited~\cite{Choi2019VisualizingFT, Liu2019DataEF}. ChartOCR \cite{ChartOCR} is another automated system that extracts data from real-world charts with better accuracy but it only predicts the raw data values of marks without extracting visual encodings. Other works focused on \change{using combinations of image processing techniques \cite{chartVi}}, 
extracting visual encodings \cite{poco2017reverse}, or recovering data from map charts \cite{poco2018extracting} and  scatterplots \cite{cliche2017scatteract}. 

Harper et al. \cite{Harper2014, Harper2018} developed a technique to extract underlying data from D3 visualizations. Extracting data and visual encodings is relatively easier for D3 charts because the developer can bind the data within the Document Object Model (DOM). Therefore, we applied the D3 deconstructor in our initial work. As deconstructing charts in bitmap image format becomes more accurate they can be used in our \seechart\ tool to create accessible visualization.

\subsection{Chart Summarization}

Researchers propose various natural language generation (NLG) methods to describe how the chart encodes data and explain “insights” in textual form for understanding charts. Early works \cite{mittal-etal-1998-describing,Ferres-accessibility-2013} followed a planning-based architecture \cite{reiter2007architecture} and used templates to generate texts. These systems only explain how to read the chart and do not convey key insights from the chart. Commercial systems like Quill\cite{quill} and Wordsmith\cite{wordsmith} and research prototypes Voder\cite{srinivasan2018augmenting} and Datasite\cite{cui2019datasite} computed key statistics (e.g., extrema, outliers) to present facts from charts. \change{Demir et al. \cite{demir-etal-2012-summarizing} used Bayesian Inference System \cite{automatedUnderstanding} to generate the summary 
from a bar chart image. The SIGHT system \cite{accessToMultiModalArticles} also used bayesian network to identify the intended message of a chart.} ChartVi \cite{chartVi} generates summaries by computing statistics using image-processing based extraction of textual information from chart images.  Datashot computes a variety of data facts covering statistics and data trends from data tables~\cite{wang2019datashot}.

One notable limitation of the abovementioned approaches is that the sentences generated using pre-defined templates may lack variations in terms of syntactic structure and lexical choices compared to data-driven models. Therefore some recent works applied data-driven deep learning models based on transformer \cite{obeid, chart-to-text} or LSTM-based encoder-decoder architecture \cite{Andrea-carenini}. While some of these works managed to generate very fluent summaries, they also suffered from
hallucinations and factual errors as well as difficulties in correctly explaining complex patterns and trends in charts. In our needfinding studies with users, we learned that people prefer factually accurate statements even if they may lack variations over incorrect facts with fluent texts. Therefore, we developed a template-based approach for chart summarization.


\section{Needfinding Interviews}
In order to understand how people with blindness attempt to read visualizations and how future tools could support their tasks effectively, we conducted a needfinding study. \final{While recent work on accessible visualizations ~\cite{jung2021communicating, HowAccessible, Kim2021, accessibile-viz} focus on understanding some key challenges and requirements of blind users while exploring visualizations, they did not focus on identifying specific design requirements for an online tool that introduces interactive accessible features for visualizations such as filtering, selection, and dynamic chart summary generation. Therefore, we conducted the needfinding interviews to further understand more specific design requirements of an accessible interactive visualization tool.}  In particular, we focused on understanding the techniques and practices most users follow and the challenges they face while understanding visualizations present on websites, and the features they would like to have in an accessible visualization tool. \change{We aimed to learn how the tools and approaches resulting from previous research have contributed to the understanding of visualizations by users with blindness. We also wanted to get early feedback and suggestions about the prototype of \seechart\ during the needfinding interviews.}


\subsection{Participants}

For the needfinding study, we recruited 15 participants (9 female, 6 male, age range 45-55 years) who identified themselves as blind \change{since birth}. In order to find a diverse range of population, we shared our study-advertisement on social media platforms (Twitter, Facebook, and LinkedIn) and also engaged with several local and national non-profit organizations for blind populations including the Canadian Council of the Blind \cite{ccb}, Canadian National Institute for the Blind \cite{cnib}, and Fighting Blindness Canada \cite{fbc}. The participants were proficient with screen-reader tools such as JAWS and NVDA for browsing Web. Seven of them occasionally encounter charts and graphs (several times a month) on websites and articles while others encounter them less frequently.


\subsection{Procedure}

\begin{figure}[t!]
\centering
  \includegraphics[width=\linewidth]{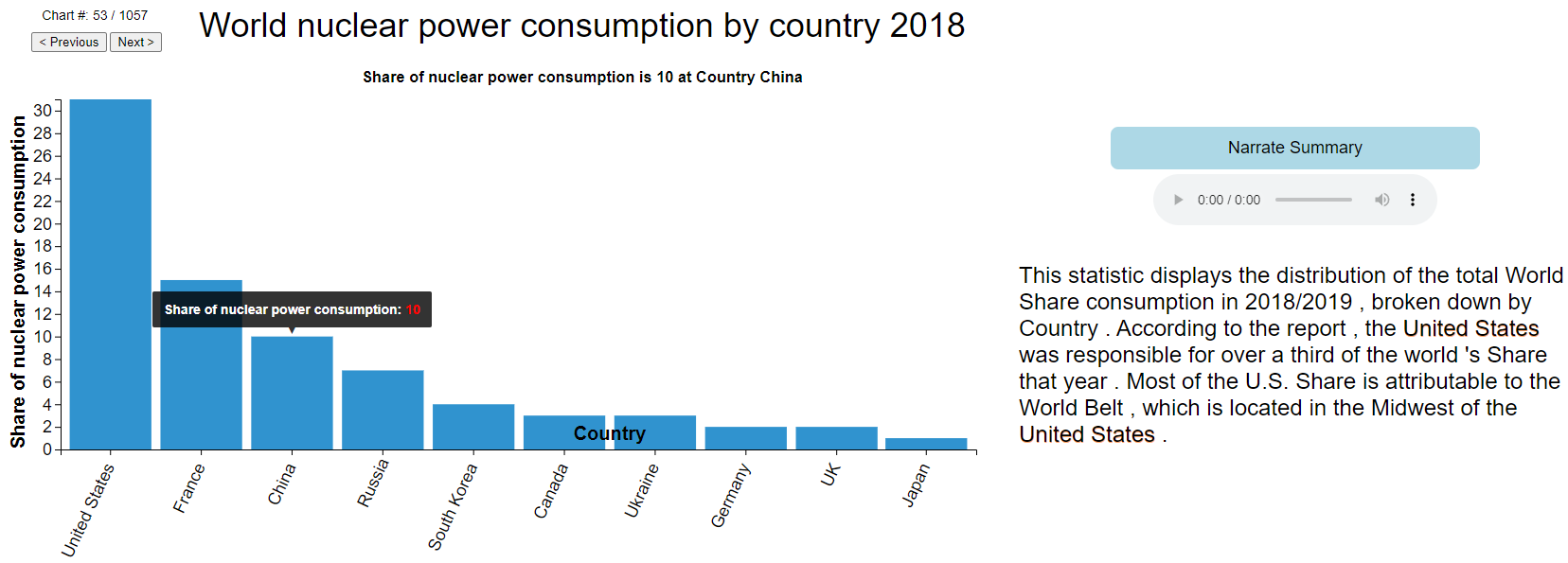}
  \caption{Mockup user prototype of \seechart\ interface.
  }
  \label{fig:prototype}
\end{figure}

During the needfinding study, we asked the participants open-ended questions regarding their \change{experiences,} challenges, needs, and requirements for reading visualizations. We started by asking ``\textit{How do you try to understand a chart that appears in a document or a web page?} and ``\textit{How important do you think it is for you to understand data visualizations?}". We also wanted to know if they find the 
alt texts helpful to understand visualizations. Finally, we asked them if they use any specific computer-based tool to explore data visualizations and what features they would like to have in an accessible tool for  visualizations. \change{Informed by these interviews, we designed a mockup user prototype of \seechart\ tool (shown in Figure \ref{fig:prototype}) 
to further comprehend the technical requirements in another course of follow-up interviews with each participant and received their feedback. The prototype was equipped with several sample charts with simple user interactions (e.g., reading data values of a chart by pressing keyboard arrow keys and hearing descriptions of the chart). 
We hosted the tool on a website and shared the web-link with the participants.
They used the tool by navigating to the link on their own computers.} 

The semi-structured interviews were conducted over video conferencing and each session lasted 40-45 minutes. As compensation, each participant was gifted 20\$. We audio-recorded interviews and transcribed them for analysis. \change{In particular, we used a spreadsheet application to thematically analyze the interview transcripts and notes taken during the interviews using the open coding method \cite{khandkar2009open}. The key themes from our analysis are presented in the following section.} 

\subsection{Findings}

We now present the primary themes found from our analysis along with the user requirements (UR) we identified from the analysis to guide the design of our \seechart\ system. 

\subsubsection{Inadequate information with the alt texts}

\change{Users with blindness generally} rely on screen-reader tools to read a web page by going through tags (e.g., \texttt{<h1>, <p>} etc.) and they control the navigation and pace according to their needs. Whenever the website provides alternative texts for visualizations using \texttt{alt} tags, \change{users} listen to \change{those texts} to make sense of the visualizations. Most participants reported that usually alt texts for charts are very generic (summarizing the topic) and do not provide important information like chart type, visual encoding, or any insights from the chart. This finding is consistent with the study from Jung et al. ~\cite{jung2021communicating} who reported that the majority of alt texts in their collected corpus do not report important information such as chart type and data trends. Participants suggested that alt texts should provide important information such as chart types, and the visual encoding structure (e.g. what each axis represents), and summarize key insights covering patterns, trends, and outliers from the visualization (\textit{UR-1}). 

Eight participants mentioned that there are many occasions where the web page presents the visualization in bitmap image format without providing any alt text or other textual summaries. In such cases, they may not be even aware of the existence of a visualization and skip it inadvertently. They suggested that an assistive tool should notify users of the presence of a chart so that they can delve into the chart (\textit{UR-2}). \change{P03 said, ``\textit{Most of the time if it's a chart, there's no way to interpret it so I just have to ignore it even if it sounds interesting to me. Overall, the situation is so upsetting that I have stopped expecting to learn useful information that is laid out in a chart. It really makes me feel like I am being deprived of publicly available information because I cannot see.}''}

\subsubsection{Chart data remains out of reach}
\change{Eleven of the participants indicated that they frequently visit websites containing charts. However, underlying data tables are rarely provided with online charts. While sometimes the alt text may mention extremum data values, they may not mention other data points. As a result, a large subset of the data points from the chart are left in darkness, and users with blindness do not have a way to drill down to these data points. Similarly, other accessibility techniques like data sonification are also limited in revealing the precise raw data values~\cite{jung2021communicating} and prone to cause cognitive overload ~\cite{sakhardande2019comparing}. P11 who used the data sonification features with SAS mentioned the challenges in discriminating between two tones that indicate two different data values of a pie chart. She said, ``\textit{SAS makes different tones for different points as it goes up or lower but it does not give me specifics. I don't know what 25\% sounds like.}"}

\change{
 While most charts on Web do not provide access to the underlying data table, when they do, participants try to navigate through the data table to comprehend the chart data. Unlike sighted people who can visually perceive the whole chart to get a quick overview of that chart  visually impaired people attempt to conceptualize a chart using table navigation technique where columns represent the data attributes encoded in the chart. They use keyboard arrow keys to navigate through the table by first reading through the column headers and then reading through individual data values in each row. 
According to P05, ``\textit{If the data is given on a table underneath the chart, I would use JAWS key commands to navigate the table. Then I would read the heading first and use the arrow keys to navigate the columns. Otherwise, I get nothing.}" Since the users have formed their mental model in accessing raw data using table navigation techniques, it is important to provide a positive transfer effect while designing the navigation for chart data (\textit{UR-3}). Feedback from our participants supports the findings from previous research \cite{understandingScreenReader}, \cite{ramloll2001using}, \cite{stockman2005interactive} as tables do not provide an overview of data and screen-reader users find it difficult to keep track of cells and columns. P05 also mentioned ``\textit{For large tables, I find it extremely hard to navigate through the data and remember them.''}}



\subsubsection{Needs for chart data filtering}
\change{Three participants who encounter charts occasionally, mentioned that they often need to filter data and extract partial information from a big chart. P07 said ``\textit{A sighted person would be able to visually perceive the whole chart to get a quick overview. But I do not get that privilege. For complex charts, I ask my colleagues to assist me with getting interesting information as there is no way of data filtering.}" In this context, an assistive tool should encourage this aspiration for exploring chart data by allowing users to select a portion of the chart that they are interested in and summarize such portion of dataset and drill down to detailed values on demand (\textit{UR-4}).}

\subsubsection{Needs for interactive features}

When we asked questions specifically about what participants would expect from a future tool for accessible visualization, they suggested to introduce interactive techniques that are consistent with their currently used tools. Users with blindness frequently use keyboard shortcuts for website navigation using screen-reader tools and a new tool should introduce interactive features that are learnable and consistent with their current forms of interactions with Web (\textit{UR-5}).

\change{During the exploration of our initial prototype, participants provided some useful suggestions for improving the interactive features.
Several participants mentioned that  controlling the information granularity in the chart summary is important to them. Often the descriptions of the visualizations are too long for a user to process as continuously listening to a long summary involving numbers can lead to cognitive overload. Therefore, the user should be able to control the length of the chart summary, hear each sentence individually, and replay a sentence (\textit{UR-6}). P03 said ``\textit{The tool plays the chart summary without any break and it is very difficult to concentrate on the numbers and labels. If there was a way to hear the description line-by-line or word-by-word, I am sure it would be very much advantageous.}" The prototype was not providing any navigational cue while participants were reading chart data values using arrow keys. It was puzzling the participants as seven of them found it very difficult to grasp where in the chart they were at and they suggested providing them with some cues so that they get an idea of their current position within the chart (\textit{UR-7}) (e.g., whether they are at the beginning or end of the chart).}



\begin{figure}[t!]
\centering
  \includegraphics[width=\linewidth]{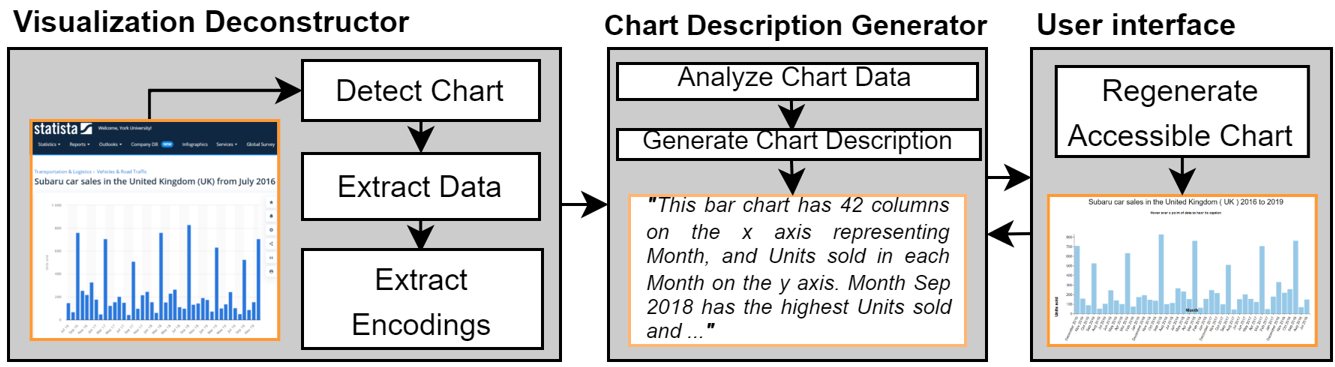}
  \caption{An overview of the \seechart\ system.
  }
  \label{fig:Pipeline}
\end{figure}

\subsection{Design Principles}
Based on our needfinding study we have identified the following design principles (DP) which formed the basis of the  \seechart\ system. Each of them was derived from one or more user requirements, as follows:
\begin{enumerate}
  \item \textbf{Give users access to data with intuitive navigational cues:} The tool should allow users to navigate through the data points in the chart via arrow keys and play each data value along with the related data attribute. The tool should also provide a positional awareness (e.g., beginning, middle, or end of the chart) as they traverse through the chart (to address \textit{UR-3, UR-7}). 
  
  \item \textbf{Provide chart details on demand:} When a chart is detected on a web page, the tool should notify the user. If the user is interested in further exploring the chart, more detailed information should be accessible through keyboard shortcuts (to address \textit{UR-1, UR-2}). 
  
  \item \textbf{Provide interactive selection of interesting area of chart:} The tool should allow users to select a range of data points following the same way they select multiple cells of a table through keyboard and generate a summary of data points based on the selected data points (to address \textit{UR-4}). Users should get feedback as they select the data points and the description should be generated in real-time.
  
  \item \textbf{Provide more control for adjusting the information:} The tool should provide a concise  textual summary of the chart with what the chart represents and what are some key insights in the chart with appropriate punctuations. The user should be able to listen to the summary sequentially and pause/skip sentences whenever needed to comprehend the information easily without getting cognitive overload. Finally, the user should be able to choose the desired length of description (addresses \textit{UR-6}).

  
 \item \textbf{Keep interactions learnable and consistent:} To enhance learnability, the interface features of the tool should be intuitive and follow familiar shortcuts. Blind users mostly rely on keyboard shortcuts while browsing Web and they use a number of shortcuts that are assigned to screen-reader tools. Introducing many additional shortcuts would be difficult to remember. Therefore, it is important to carefully choose a small number of shortcuts to interact with visualizations  (to address \textit{UR-5}).   
  
\end{enumerate}

\section{The SeeChart System}

\begin{figure}[t!]
\centering
  \includegraphics[width=0.9\linewidth]{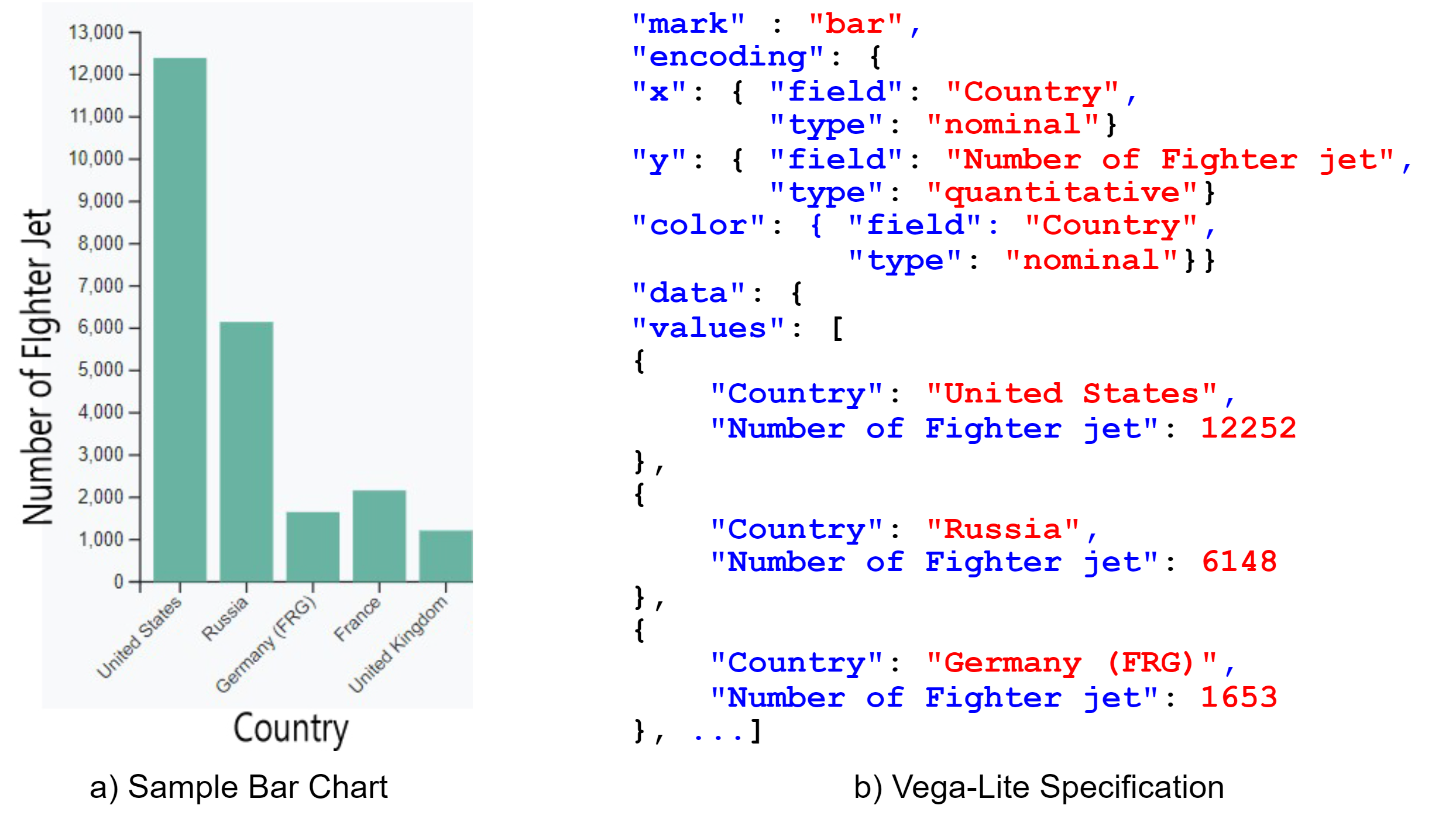}
  \caption{A D3 bar chart (a). D3 deconstructor extracts data fields ``Country" and ``Number of Fighter Jet" of data types ``nominal" and ``quantitative" respectively. Also, the axes, including the data field they represent are extracted. The deconstructed representation for the bar chart is a Vega-Lite specification (b)}
  \label{fig:decon}
\end{figure}

Our system has three main modules. Given a web page with an SVG visualization loaded in the Google Chrome browser, the \textit{Visualization deconstructor} analyzes the SVG representation of the chart to recover the data, visual attributes of graphical marks, and visual encodings that describe the relationship between them. The \textit{chart description generator} then takes the data, marks, and visual encodings as input and automatically generates the chart summary as output. Finally, the \textit{user interface} reconstructs the visualization with summary descriptions as well as provides several interactive features for selection, navigation, and browsing through the chart. \final{Figure \ref{fig:Pipeline} shows the pipeline of \seechart\ system.}

\subsection{Visualization Deconstructor}
We developed the visualization deconstructor as an extension for Google Chrome browser which injects code into the JavaScript environment for the given web page. It first detects a visualization in SVG format and then extracts the underlying data and visual encodings. We have primarily focused on deconstructing visualizations created using the D3.js, a JavaScript library \cite{d3} which has become very popular in constructing web-based visualizations. 
We utilized the D3 deconstructor tool \cite{Harper2014, hoque-vis-search-2019} which analyzes the relationships between data values and the values of each mark attribute to recover the encodings that map between them. Figure \ref{fig:decon} shows an input bar chart and the resulting deconstruction output in Vega-Lite format. In addition to deconstructing D3 visualizations, we also deconstruct  visualizations created with the highcharts library\cite{highchart} where we automatically analyze the content of the SVG to extract the data and visual encodings of charts. 

\change{Deconstructing visualizations from highcharts\cite{highchart} was relatively straightforward because charts are generated with declarative specifications and the library internally assigns specific class names for chart elements in the generated SVG chart. For example, when data values of attributes appear as text labels in the chart they are grouped together under \texttt{<g>} with \texttt{class = highcharts-data-labels}. Similarly, we retrieve other chart information such as axes labels using specific class names (e.g., \texttt{highcharts-yaxis}). In order to construct the underlying data table of the chart, we simply iterate through each of this \texttt{<g>} elements with \texttt{class = "highcharts-data-labels"}  to find data values of each data attribute encoded in the given chart. When data-labels are absent we use the bounding box information of data encoding marks which are children of \texttt{<g class="highcharts-series-group"}. We then obtain the scale information based on the axis labels and tick marks  to recover the data values for these data encoding marks.
We have deconstructed a variety of bar charts, line charts and pie charts created by the highcharts library.} 




\subsection{Natural Language Description Generator}
This module automatically summarizes the given chart explaining how the chart encodes data and what are some key takeaways from the chart. In order to learn what should be the content (\textit{what to say}) and structure (\textit{how to say it}), we analyzed an existing chart description corpus for screen-reader users\cite{accessibile-viz}. Informed by the corpus study, we generate chart summaries following the natural language generation (NLG) architecture which has been widely adopted for generating text summaries from numeric input data \cite{reiter2007architecture}. We discuss the corpus study followed by our chart summarization method below.



\subsubsection{Chart Summary Corpus Study} In order to understand what information is important and should be included in the chart summary, we rely on the recent study with blind people by Laundgard et al. \cite{accessibile-viz}. They curated 50 visualizations comprised of 18 bar charts, 21 line charts, and 11 scatter plots from MassVis dataset \cite{borkin2013makes, borkin2015beyond} and were able to gather a corpus of 582 summaries consisting of 2,016 sentences. Based on the semantic content, the sentences were categorized into four levels: (1) elemental and encoded properties such as chart type, title, axis labels; (2) statistical concepts and relations such as extremas, comparisons, and correlation; (3) perceptual and cognitive Phenomena such as trends and patterns; and (4) contextual and domain-specific insights. In our analysis, we mainly focused on understanding what information is frequently reported in Level 2 and 3 types of content that describe statistics as well as patterns, trends, and outliers in the given chart. We did not focus on Level 4 because generating contextual or domain-specific description that is not given in the chart is extremely challenging and are not of primary interest to blind people\cite{accessibile-viz}. In contrast, the Level 1 content type, which describes the chart type, title, axes labels, range, and encoding channels, is relatively straightforward to generate, therefore we manually build templates for this level.

For our analysis, two authors from our research team independently went through all level 2 and 3 types of summary contents and came up with a categorization of information presented in these sentences (e.g., \textit{extrema}, \textit{comparisons[relative]},  \textit{trend [global]]}, where each category represents a function along with possible parameters). The two annotators resolved each inconsistency by explaining their categories and  then worked together to reach a final consensus on what should be the category for each sentence. \final{Based on this analysis of 464 sentences about bar charts and 818 sentences about line charts}, we identified the frequently reported statistics and categorized each sentence into one of those categories. The `others' represent the category that does not fall into any of the identified major categories (e.g., reporting a random data point in the chart). 


We ranked the statistics and trend categories based on their frequency of occurrence as shown in table \ref{tab:barStats}. Among the 272 distinct summaries of 11 single bar charts, majority of them highlighted on \textit{extremum} (min/max) values, relative or absolute \textit{comparison} between data points, \textit{order} of data values, \textit{derived} values (e.g., average, total), and other statistical information. Almost similar kinds of categories were found in 192 summaries of 6 grouped bar charts. Line charts generally depict changes in trends over a period of time and this was no different as about 60\% of 818 summaries of 10 single and 11 multi-series line charts describe positive or negative \textit{trends}. Other notable statistics were about \textit{extrema}, \textit{comparison}, \textit{order}, etc.

\begin{table}[t!]
\centering
\caption{Ranked list of identified statistics and data trends according to their occurrence ratio.}
\label{tab:barStats}
\scalebox{1}{
\begin{tabular}{|cl|cl|}
\hline
\multicolumn{2}{|c|}{\textbf{Single Bar}}                                          & \multicolumn{2}{c|}{\textbf{Grouped Bar}}                                         \\ \hline
\multicolumn{1}{|c|}{\textit{Stat Category}} & \multicolumn{1}{c|}{\textit{Ratio}} & \multicolumn{1}{c|}{\textit{Stat Category}} & \multicolumn{1}{c|}{\textit{Ratio}} \\ \hline
\multicolumn{1}{|r|}{\begin{tabular}[c]{@{}r@{}}Extrema {[}min/max{]} \\ Comparison {[}relative{]} \\ Order {[}rank{]} \\ Trend \\ Computed derived value {[}avg/sum{]} \\ Order {[}rank/position{]} \\ Comparison {[}absolute{]} \\ Others\end{tabular}} &
  \begin{tabular}[c]{@{}l@{}}57 \%\\ 12 \%\\ 8 \%\\ 7 \%\\ 6 \%\\ 4 \%\\ 2 \%\\ 4 \%\end{tabular} &
  \multicolumn{1}{r|}{\begin{tabular}[c]{@{}r@{}}Global Extrema {[}min/max{]}\\ Trend {[}global{]}\\ Comparison {[}relative{]}\\ Extrema {[}max difference{]}\\ Trend {[}local{]}\\ Order {[}rank/position{]}\\ Local Extrema {[}min/max{]}\\ Others\end{tabular}} &
  \begin{tabular}[c]{@{}l@{}}36 \%\\ 18 \%\\ 13 \%\\ 13 \%\\ 7 \%\\ 4 \%\\ 2 \%\\ 7 \%\end{tabular} \\ \hline
\multicolumn{2}{|c|}{\textbf{Single Line}}                                         & \multicolumn{2}{c|}{\textbf{Multi Line}}                                          \\ \hline
\multicolumn{1}{|c|}{\textit{Stat Category}} & \multicolumn{1}{c|}{\textit{Ratio}} & \multicolumn{1}{c|}{\textit{Stat Category}} & \multicolumn{1}{c|}{\textit{Ratio}} \\ \hline
\multicolumn{1}{|r|}{\begin{tabular}[c]{@{}r@{}}Trend\\ Extrema {[}min/max{]}\\ Comparison {[}absolute/relative{]}\\ Extrema {[}difference{]}\\ Others\end{tabular}} &
  \begin{tabular}[c]{@{}l@{}}62 \%\\ 22 \%\\ 10 \%\\ 4 \%\\ 2 \%\end{tabular} &
  \multicolumn{1}{r|}{\begin{tabular}[c]{@{}r@{}}Global Trend\\ Order {[}position/rank{]}\\ Extrema {[}min/max{]}\\ Extrema {[}local/global{]}\\ Comparison {[}absolute/relative{]}\\ Others\end{tabular}} &
  \begin{tabular}[c]{@{}l@{}}48 \%\\ 24 \%\\ 13 \%\\ 5 \%\\ 2 \%\\ 8 \%\end{tabular} \\ \hline
\end{tabular}}
\end{table}

\begin{figure}[t!]
\centering
  \includegraphics[width=.9\columnwidth]{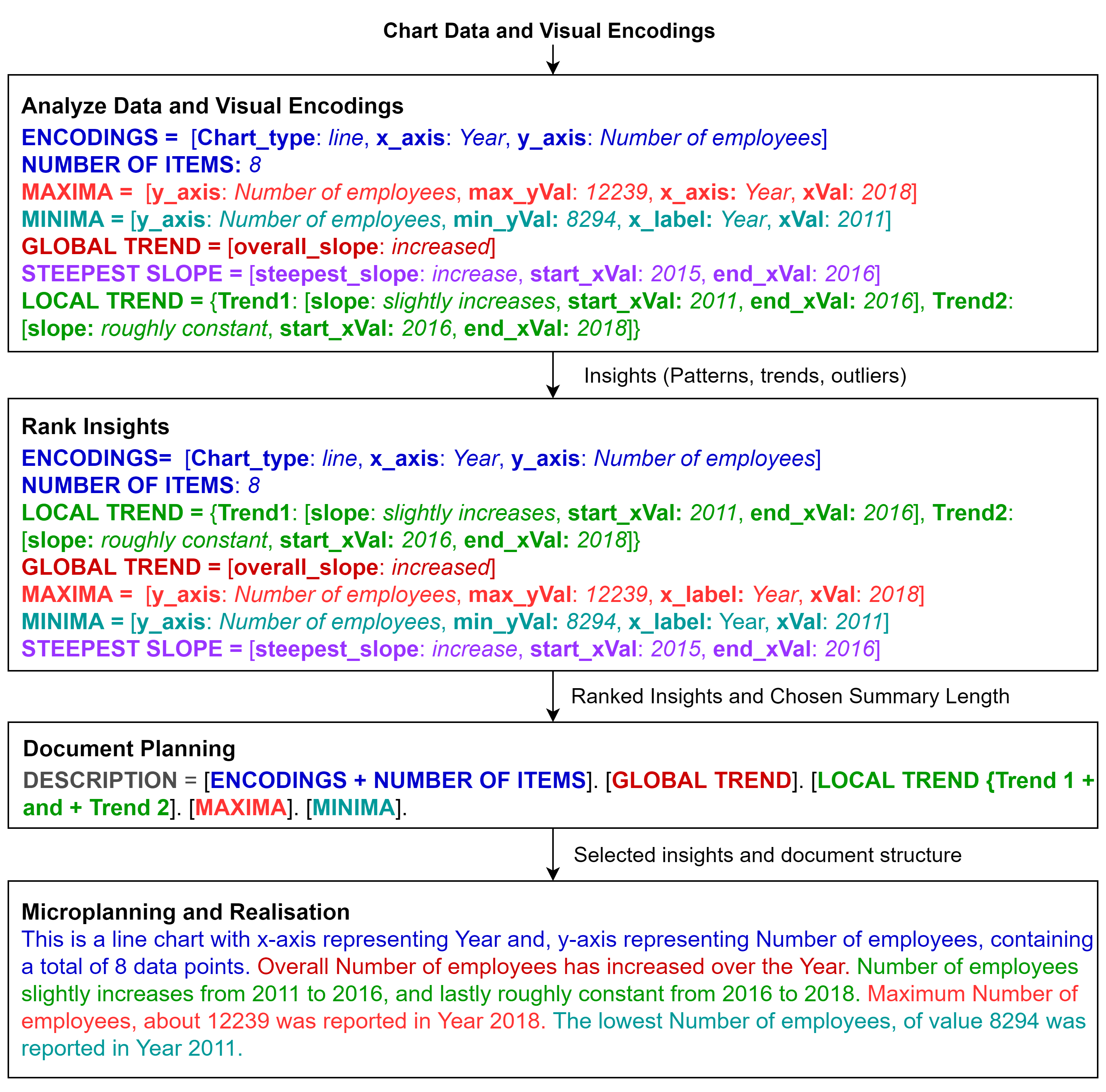}
  \caption{Architecture of template-based description generator.}
  \label{fig:architecture}
\end{figure}

\begin{figure*}[t!] 
\begin{spacing}{.5}
\selectfont\centering 
\scalebox{1}{\begin{tabular}{p{4.3cm} | p{2.5cm} | p{3cm} | p{3.9cm}}     
\toprule
        \raisebox{-.9\height}{\includegraphics[width=4.4cm, height=2.3cm]{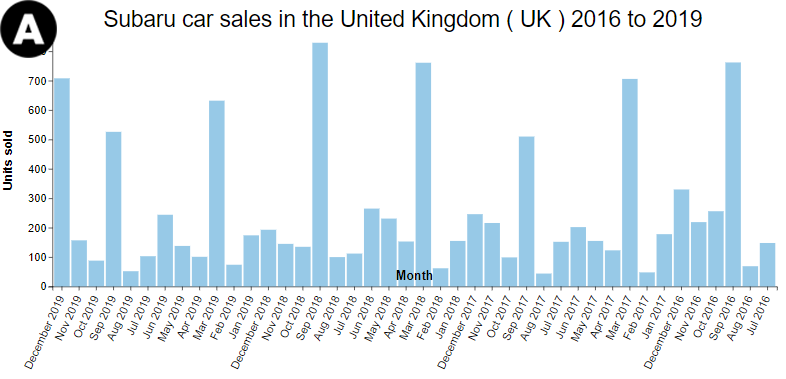}}
        &
        {\tiny \textbf{Human-Written:} This statistic shows the monthly amount of cars sold by Subaru in the United Kingdom (UK) between July 2016 and December 2019. Peaks in registration numbers were recorded in March and September of each year, which was due to the issuing of license plates by the Driver \& Vehicle Licensing Agency in those months. In December 2019, 708 new Subaru cars had been sold.}
        &
        {\tiny \textbf{Template-based (Moderate length):} This is a bar chart. It shows Units sold for 42 number of Months. The maximum Units sold 829 is found at Month Sep 2018. The minimum is found at Aug 2017 where Units sold is 44. The highest Month Sep 2018 has 785 more Units sold than the lowest Month Aug 2017.}
        &
        {\tiny \textbf{Template-based (Long length):} This bar chart has 42 columns on the x axis representing Month, and Units sold in each Month on the y axis. Month Sep 2018 has the highest Units sold and Aug 2017 has the lowest Units sold. The second highest Units sold is in the Month. There is a difference of 785 between the maximum Month Sep 2018 and minimum Month Aug 2017. Month Sep 2018 is higher than any other Months with value 829, followed by , and Mar 2018. Down to Month Aug 2017 with the lowest value 44. Over the past 42 Months, the Units sold has declined. Month Jan 2018, and May 2017 have the same Units sold. The average Units sold in all 42 Months is 252.4. Summing up the values of all Months, we get total 10601.}
        \\  
\midrule 
        \raisebox{-.9\height}{\includegraphics[width=4.4cm,height=2.4cm]{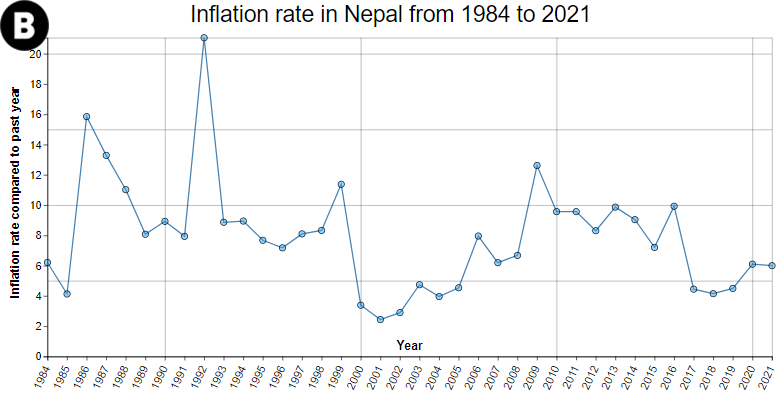}}
        & 
        {\tiny \textbf{Human-Written:} In 2018, the average inflation rate in Nepal was at 4.15 percent, a slight drop compared to the previous year. The inflation rate is calculated using the price increase of a defined product basket. This product basket contains products and services on which the average consumer spends money throughout the year.}
        &
        {\tiny \textbf{Template-based (Moderate length):} This is a line chart with an x axis representing Year and a y axis representing Inflation rate compared to previous year, containing a total of 41 data points. In general Inflation rate compared to previous year has fallen over the Year. The most drastic increase took place within the Year 1991 and 1992. Maximum Inflation rate compared to previous year, about 21.06 was reported at Year 1992 The lowest Inflation rate compared to previous year, of value 2.44 was reported at Year 2001.}
        &
        {\tiny \textbf{Template-based (Long length):} This is a line chart where the x axis denotes Year and a y axis denotes Inflation rate compared to previous year. In total the number of data points it has is 41. Overall Inflation rate compared to previous year has decreased over the Year. The chart in general has a zig-zag shape. The most drastic increase took place within the Year 1991 and 1992. Maximum Inflation rate compared to previous year, about 21.06 was reported at Year 1992 The lowest Inflation rate compared to previous year, of value 2.44 was reported at Year 2001 The line rapidly increased by 283.78\% from Year 1985 to 1986,increased by 165.24\% from Year 1991 to 1992,decreased by 70.21\% from Year 1999 to 2000,increased by 75.33\% from Year 2005 to 2006, and lastly, increased by 88.92\% from Year 2008 to 2009.}
        \\  
\midrule  
        
        \raisebox{-.9\height}{\includegraphics[width=4.4cm,height=2.4cm]{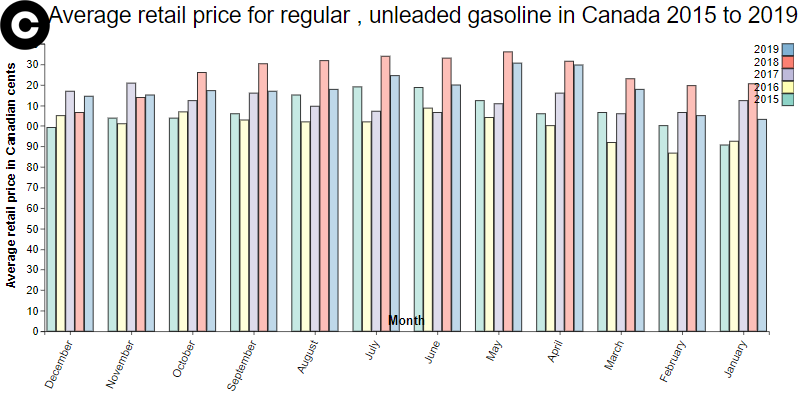}}
        &
        {\tiny \textbf{Human-Written:} The average retail price for regular, unleaded gasoline at self-service stations in Canada was 114.5 Canadian cents per liter in December 2019. Prices hit a high of 1.3 dollars per liter in May 2019. The average price in 2018 was around 1.26 dollars per liter, a substantial increase from an average of 1.07 dollars in 2015.}
        &
        {\tiny \textbf{Template-based (Moderate length):} This is a grouped bar chart showing Average retail price in Canadian cents per liter on the Y-axis throughout 12 Months for 2015, 2016, 2017, 2018, and 2019 on the X-axis. Over the 12 Months, the Average retail price in Canadian cents per liter for both 2016 and 2015 have reduced. For Month May, 2018 had the highest Average retail price in Canadian cents per liter 1359 among the 5 groups and in Feb, 2016 had the lowest Average retail price in Canadian cents per liter 867. Month Jun and Feb had the biggest gap of 219 between the highest and lowest Average retail price in Canadian cents per liter found for 2016.}
        &
        {\tiny \textbf{Template-based (Long length):} This is a grouped bar chart showing Average retail price in Canadian cents per liter on the Y-axis throughout 12 Months for 2015, 2016, 2017, 2018, and 2019 on the X-axis. All through the Months, similar trend was observed for 2016 and 2015. In both cases, the Average retail price in Canadian cents per liter have reduced. Out of all 5 groups, 2018 had the highest Average retail price in Canadian cents per liter for Month May and 2016 had the lowest Average retail price in Canadian cents per liter for Month Feb. The maximum Average retail price in Canadian cents per liter for 2016 that was found in Month Jun was 1.25 times larger than the minimum Average retail price in Canadian cents per liter observed in Month Feb. Among all 5 groups, 2018 had the gap of 295 between the maximum and minimum Average retail price in Canadian cents per liter observed in Month May and Dec respectively. The second highest Average retail price in Canadian cents per liter 1191 was observed for 2018 in Month May. On average, the Month May had the highest Average retail price in Canadian cents per liter for all 5 groups 2015, 2016, 2017, 2018, and 2019. Whereas Feb had the lowest average Average retail price in Canadian cents per liter.}
        \\
\midrule  
        
        \raisebox{-.9\height}{\includegraphics[width=4.4cm,height=2.4cm]{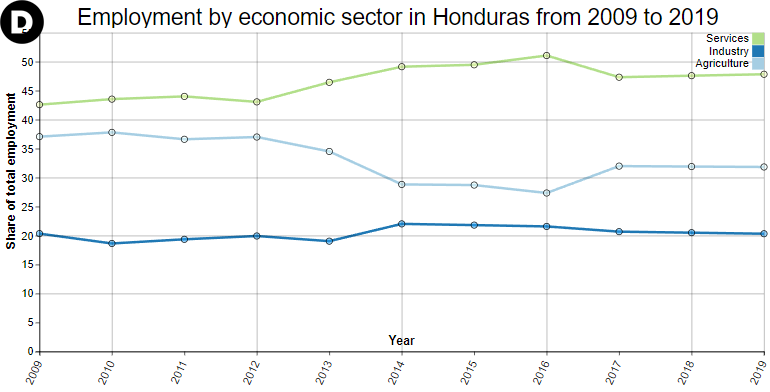}}
        &
        {\tiny \textbf{Human-Written:} The statistic shows the distribution of employment in Honduras by economic sector from 2009 to 2019. In 2019, 31.84 percent of the employees in Honduras were active in the agricultural sector, 20.32 percent in industry and 47.84 percent in the service sector.}
        &
        {\tiny \textbf{Template-based (Moderate length):} This is a multi line chart with 3 lines representing Agriculture, Industry, and Services. The y axis denotes Share of total employment and the x axis denotes Year. Overall Services is gradually increasing, Agriculture is gradually decreasing, Industry is roughly constant, throughout the Year. Throughout the period, Services generally had the highest Share of total employment relative to others with an average of 46.56, and it reached its maximum at 2016 with a value of 51.07. After Services, Agriculture overall has the second highest values , with a mean value of 33.0, peaking at 2010. Industry mostly had the least Share of total employment with a mean value of 20.38, which peaked at 2014 with a value of 22.02.}
        &
        {\tiny \textbf{Template-based (Long length):} This is a multi line chart with 3 lines representing Agriculture, Industry, and Services. The y axis denotes Share of total employment and the x axis denotes Year. Overall Services is gradually increasing, Agriculture is gradually decreasing, Industry is roughly constant, throughout the Year. The lines ordered according to average values of Share of total employment in descending order is: 1. Services, 2. Agriculture, and lastly, 3. Industry. Overall across the Year, Services mostly maintained the highest Share of total employment when compared to others with a mean value of 46.56, and it peaked at 2016. Followed by Agriculture which ranks as the second topmost line , with an average of 33.0 Share of total employment,reaching its highest point at 2010 with a value of 37.8. The bottommost line, Industry, has a mean of 20.38, and peaked at 2014. The minimum Share of total employment about 18.64, occurred at 2010 by Industry.}
        \\        
\bottomrule  
    \end{tabular}
    }
    \caption{Sample charts from Statista in the first column and their human-written summaries as well as summaries generated by our template-based algorithm (both moderate and long lengths) are shown in next three columns (We encourage readers to zoom in to see chart details and summaries).}
    
    \label{tab:examples}
\end{spacing}
\end{figure*}

\subsubsection{Chart Summarization Method} To generate summaries from chart data, we followed the natural language generation (NLG) architecture proposed by Reiter which has been widely adopted for generating text summaries from numeric input data \cite{reiter2007architecture}. Following  this data-to-text generation architecture, our NLG method works according to the following stages as shown in Figure~\ref{fig:architecture}. 

\noindent \textbf{\textit{Stage 1: Analyze data and visual encodings.}} The goal at this stage is to detect basic statistics and patterns in the numeric input data such as maxima, minima, overall trend in the chart, or a short-term trend within a range of data items (e.g., spikes). Some examples of the resulting statistics and patterns are shown in Figure~\ref{fig:architecture} (top block). For characterizing trends, we follow Hullman et al. for identifying visually salient points in line chart \cite{hullman2013contextifier}.
However, their method included some domain-specific features (e.g., stock price and volume traded) to calculate visual saliency of each data point. We simplify the approach by considering the variable that  captures the change (slope) over data point in the previous time point by computing the absolute distance between the value for the current date point and the value for the previous time point. We then normalize the value of change by dividing by the maximum change value across the series. Based on the calculated slopes, we compute the global trend over the whole chart as well as local trends over different portions of the chart. We also used the extreme values in each series of the chart and the change of magnitude and the direction of a slope (e.g., from increase to decrease) at different points to determine the range of a local trend. 

\noindent \textbf{\textit{Stage 2: Rank insights.}} At this stage, the goal is to assign importance to each of the insights and then rank them according to the importance measure. For ranking each message, we use a simple measure of how frequently the corresponding insight appears in chart summaries for that chart type.  According to  Table \ref{tab:barStats}, people tend to report extrema and comparisons between values for bar chart, whereas for line charts people focus more often on trends. Nevertheless, in the future other measures of importance 
maybe combined with this frequency information for ranking statistics.

\noindent \textbf{\textit{Stage 3: Document planning.}} In this stage, the method determines the content (which messages should be included in the final summary) and the structure of the summary (what would be the discourse structure of the text, e.g., how to combine multiple messages into a sentence and how to determine sentence boundary). We first use a set of heuristics to combine related statistics, for example, maxima and minima values into a single sentence. Similarly, multiple short-term trends are joined together in a sentence using discourse connectives such as `and', `however', `after that' etc. We then first present the introductory message containing the basic chart information and visual encoding channels  such as graph title, type of graph, label of X and Y axes which are identified as important for blind people by the study from Ferres et al. \cite{Ferres2006}. The remaining messages are presented according to their ranking. For example, when the maximum summary length is 3 sentences, we  present the intro sentence (describing chart type and visual encoding channels) followed by the top two messages identified after ranking statistics, which are \textit{extrema (min/max)} and then \textit{comparison (absolute/relative)}.

\noindent \textbf{\textit{Stage 4: Microplanning and realization.}} The last stage of the architecture generates actual text based on the content and structure determined in the previous stage. To avoid too simple and repetitive structure, we analyzed three different corpora of summary sentences for charts and curated several templates for each message type with varying syntactic structures. To find various variations of templates, we analyzed the dataset shared by Kanthara et al. in their extended work on Chart-to-Text\cite{new_c2t}. We randomly pick a different template every time a particular category of message appears to enhance naturalness and diversity in lexicons. For example, the intro message for the single bar chart shown in Figure \ref{tab:examples} (A), our tool will generate following alternative intro  messages and pick one of them randomly: \textbf{1.} ``\textit{This is a bar chart representing 42 Months in the x axis and Units sold in the y axis.}" \textbf{2.} ``\textit{This bar chart has 42 columns on the x axis representing Month, and Units sold in each Month on the y axis.}" \textbf{3.} ``\textit{This is a bar chart. It shows Units sold for 42 number of Months.}"

Similarly, for the \textit{extrema(min/max)} category, one of the following sentences would be picked:
\textbf{1.} ``\textit{The maximum car sold 829 is found in Month September 2018 and the minimum is found in August 2017 where Units sold is 44.}" 
\textbf{2.} ``\textit{The Units sold was highest in Month September 2018 and lowest in Month August 2017.}" 
\textbf{3.} ``\textit{The Units sold was appeared to be the highest in Month September 2018 and lowest in Month August 2017.}"

\textbf{Sample Summary Outputs:} Figure \ref{tab:examples} shows example charts and corresponding summaries: single bar (A), single line (B), grouped bar (C), and multi line (D) taken from Statista. The second column contains the human-written summaries of the charts that we found in Statista. Descriptions generated by our template-based algorithm (with moderate and long length) for corresponding charts are presented in last two columns. Note that \seechart\ allows user to change the description length of a chart to 3 levels: \textit{short}, \textit{moderate}, and \textit{long} length. \textit{Short} summaries are generally 3 to 4 sentences long depending on whether it is a bar chart or a line chart.

\begin{figure*}[h!]
\centering
  \includegraphics[width=.96\linewidth]{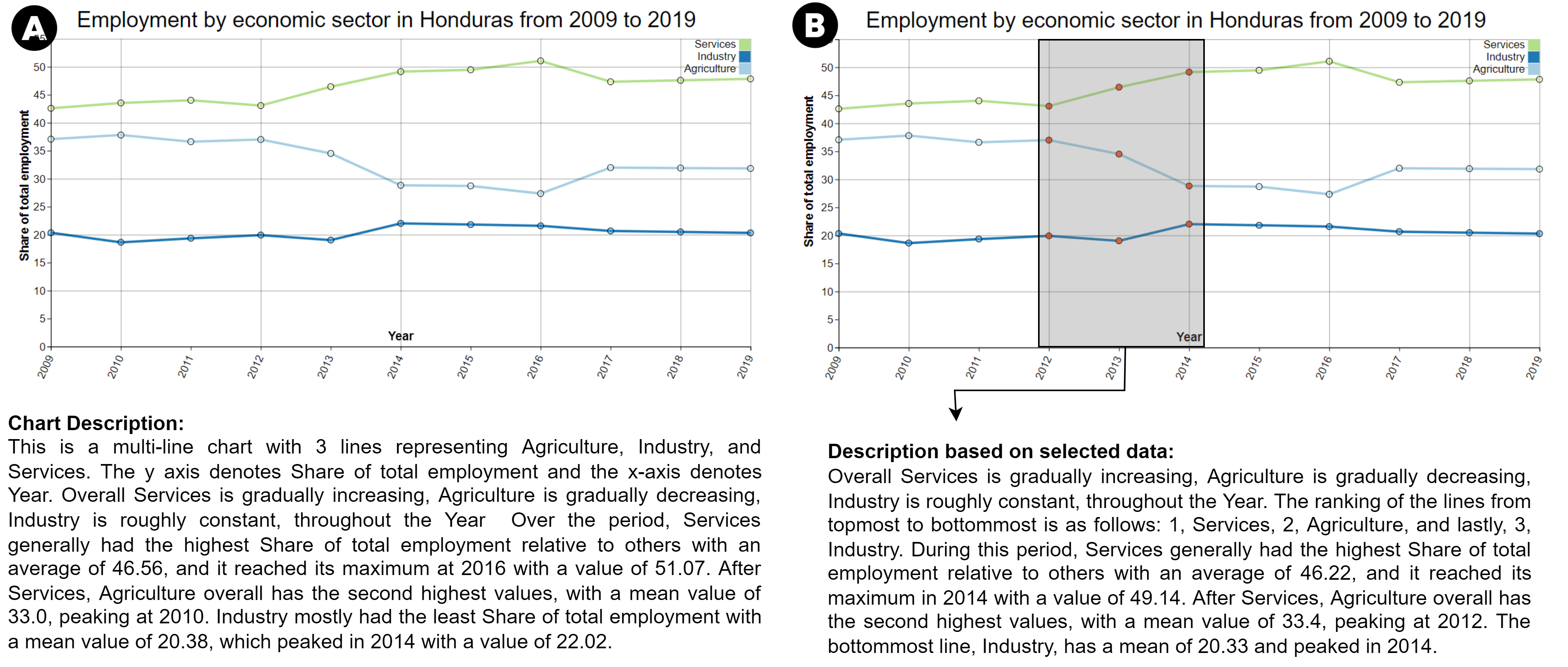}
  \caption{
  Initial summary of a line chart (A), and partial summary based on selected data points (B).}
  \label{fig:partial}
\end{figure*}

\subsection{The \seechart\ Interface}
Informed by the insights gained from needfinding interviews with our participants, we designed and developed the \seechart\ interface which takes a chart appeared on a webpage and regenerates an interactive version of it with audio description of the chart as well as features for selecting and navigating through charts. During our design process, we involve three participants from the needfinding interviews who interacted with our initial prototype and provided feedback. We have considered the feedback to improve the accessibility features. All these interactive features are accessible through keyboard shortcuts.

\subsubsection{Get an Audio Overview of the Chart} When \seechart\ detects a chart on a webpage, it reconstructs the chart and generates the chart description in real time. Upon successful reconstruction of the chart summary, the tool plays a \textit{chime} sound to notify the user about the presence of the chart. The sound does not intervene in any ongoing narration being played by the screen-reader tools. At this point, if the user wants to explore that chart, pressing the \texttt{SPACE} key opens the reconstructed chart in a pop-up window to let user access the chart through a set of interactive features. 

Users can get a quick overview of what the chart is about by pressing \texttt{Enter} which provides the chart title along with chart type, e.g., \textit{``This is a Bar chart.  It shows Subaru car sales in the United Kingdom ( UK ) 2016 to 2019''}. The user can access the chart summary by pressing the \texttt{Space} key. However, \change{during exploration with the \seechart\ prototype,} participants mentioned that going through the long summary at once results in cognitive burden as they need to keep track of the information and overload their working memory. To address their concern, we enable them to hear the summary sentence-by-sentence by pressing \texttt{J}, \texttt{K}, and \texttt{L} keys, where \texttt{J} and \texttt{L} play the next or previous sentence, and \texttt{K} repeats the sentence. Since the right hand of a user typically rests on the \texttt{J}, \texttt{K}, and \texttt{L} keys, we used these keys to make sure users find it easy to learn and remember (\textit{DP-5}). By default, the description is provided at a moderate length, however, users can adjust the summary length by pressing \texttt{1} (short), \texttt{2} (moderate), or \texttt{3} (long) (\textit{DP-4}).

Note that if there are multiple charts on a web page, the \seechart\ system reconstructs all of them and user can navigate between them by pressing \texttt{N} (next) and \texttt{P} (previous) keys. All audio descriptions are generated with proper punctuation so that users can comprehend them effectively (\textit{DP-5}).

\subsubsection{Access to Chart Data}


	
	
	

Users with blindness generally do not get access to chart data as most of the websites do not provide chart data in a separate table. Therefore, there is a pressing need for browsing chart data in a similar way they browse table data on Web. Participants suggested that when they encounter data tables on a webpage, they read the data of each cell by using the keyboard \texttt{arrow} keys. Therefore,  \seechart\ uses such shortcuts so that users explore a chart by pressing \textit{arrow keys} to  navigate through the data points from \change{the beginning to the end} or vice versa (\textit{DP-1}). For each data point, the interface plays the information such as the data value and the corresponding data attribute(s).
For example, for the bar chart shown in Figure \ref{fig:teaser} (E), if the user selects the first data point by pressing the \texttt{right arrow} key, the tool will say ``\textit{In airlines American, the number of passengers in millions was, 203.}". For bar and single line charts, user can only navigate to left or right by pressing the corresponding \texttt{arrow} keys. When line charts have multiple series the user can switch among different line series by pressing \texttt{up} or \texttt{down arrow} keys. \change{Similarly, in the case of a grouped bar chart, the user can move between different groups by pressing \texttt{up} or \texttt{down arrow} keys.}

The interface also provides navigational cues by informing where the user is navigating right now. For example, when the user selects the first bar from left in a vertical bar chart, the system reminds the user by saying ``\textit{You are at the beginning of the chart}" or ``\textit{This is the last point of the graph}". When the user reaches the end of the chart and presses the right arrow again, she would be redirected to the first point and again informed of the current position. \change{Similarly, when users reach the topmost or bottommost line, \seechart\ informs the user by saying ``\textit{This is the topmost line"} or ``\textit{This is the bottommost line}".} This allows users to explore a chart without getting disoriented. 


\subsubsection{Interactive Summarization}
While accessing individual points is helpful to users, they may have difficulty comprehending a large number of data points together. Sometimes they desire to select a region of interest in the chart and get the corresponding summary for that selection. \seechart\ allows users to select the data points they are interested in and provides a comprehensive description based on the selected points (\textit{DP-3}).  

The user can press the \texttt{Shift} key to activate the multi-point selection mode, then continue to select using the \texttt{Right-arrow}. We adopt this interaction because this is how we commonly perform selection in various tools, for example, selecting text using a word processing tool (\textit{DP-5}). The interface provides feedback as the user continues to select (e.g., ``\textit{Year 2012 is selected}"). Once user selects desired data points and releases the \texttt{Shift} key, \seechart\ generates the summary of the selected range of data immediately using the template-based algorithm as described in Section 4.1.  The user can reset the selected points by pressing the ``\texttt{Escape}" key. The user can also listen which points have been selected by pressing the ``\texttt{S}" key. Figure \ref{fig:partial} shows an example of partial summary generated based on selected data points: \texttt{year 2012}, \texttt{2013}, and \texttt{2014}.  Finally, users can also get a summary from discontinuous sets of data points from the chart (e.g., by selecting \texttt{2012, 2013, 2014} followed by \texttt{2018, and 2019}).

\seechart\ also supports direct manipulations to choose multiple items through lasso selection and brushing. However, 
according to our needfinding study, our participants preferred to use keyboard shortcuts rather than using mouse. Future work may explore how they may perform these interactions with touch and tactile-based interfaces. 

\subsubsection{Keyword-based Search} 
\label{voice}
\change{Based on the feedback of some participants during our main user study, we have integrated a voice search feature in the \seechart\ interface.}

\change{We have used Web Speech API\cite{webspeech} to construct the speech recognition system and user can activate the search mode by pressing the ``\texttt{F}" key and then typing the question or activating the voice search mode by pressing the ``\texttt{Q}" key.
Upon successful recognition of a word or phrase the tool initiates the search operation and looks for predefined keywords (e.g., \textit{x-axis data label}, \textit{maximum}, \textit{minimum}, \textit{trend}, etc.) in the search string, \seechart\ presents the result if found or asks user to ask the question again with some hints of what they can ask. User can also type the question after activating the search mode and press ``\texttt{Enter}" to get the answer. When the search mode is active, no other assigned shortcut keys will work unless user presses ``\texttt{ESC}" key or complete the operation by pressing ``\texttt{Enter}". 

Using this feature, users can find the corresponding y-axis value of an x-axis data label. For the chart shown in Figure \ref{fig:partial}, user can enter a query like ``\textit{What is the value of 2011?}" or simply say ``\textit{2011}" and if the provided data is accurate, the tool will reply ``\textit{We have found multiple values for Year 2011. These are, Agriculture is 36.62, Industry is 19.36, Services is 44.02.}". 
Users can learn what type of questions can be asked by checking the help menu.
}

\subsubsection{Control Audio Speed}
Screen-reader users rely on auditory feedback to communicate with computer devices. Screen-reader tools let users select faster speech rates to speed up the consumption of digital information. Previous study\cite{voice_rate} showed that, at around 278 words per minute (WPM), visually impaired users found the sweet spot between scanning information efficiently and understanding each sentence while average person speaks at 100-130 WPM\cite{avg_voice}. In \seechart, we used the \textit{Text-to-Speech} (TTS) service provided by Amazon Polly\cite{polly} to convey information from charts (e.g., the value of a data point, the description of a chart, and other commands to guide users). Since the tool primarily describes quantitative information, \change{it allows the user to toggle between different voice rates by pressing the ``\texttt{T}" button to understand the information at their own pace.}

\subsubsection{Keyboard Shortcuts}

\seechart\ comes with a list of shortcut keys for different functionalities (as shown in table \ref{tab:shortcuts}).
We tried to keep the number of keys and their combinations minimum.  In our needfinding interviews, one participant (P05) suggested: ``\textit{If you are planning to assign keyboard shortcuts, try to avoid too many key combinations as I already need to remember many shortcuts for JAWS}". \seechart\ has only one shortcut that involves multi-keys press: \texttt{Shift + Right Arrow} since this is the standard way of selecting texts (Microsoft Word) or cells (Microsoft Excel). Also, we tried to keep finger-travel distance minimum by utilizing the keys where human hands rest for frequently used functions (e.g., \texttt{J}, \texttt{K}, and \texttt{L}).

\begin{table}[h]
\centering
\caption{List of available keyboard shortcuts in \textit{SeeChart}.}
\vspace{2mm}
\label{tab:shortcuts}
\scalebox{0.9}{
\begin{tabular}{@{}cl@{}}
\toprule
\textbf{Key/s} &
  \multicolumn{1}{c}{\textbf{Functions}} \\ \midrule
\multicolumn{1}{r}{\begin{tabular}[c]{@{}r@{}}\texttt{Enter}\\ \texttt{X,Y}\\ \texttt{Space}\\ \texttt{J, L}\\ \texttt{K}\\ \texttt{F}\\ \texttt{Q}\\ \texttt{Right or Left Arrow}\\ \texttt{Up} or \texttt{Down Arrow}\\ \texttt{Shift + Right Arrow}\\ \texttt{S}\\ \texttt{Escape}\\ \texttt{1, 2, 3}\\ \texttt{T}\end{tabular}} &
  \begin{tabular}[c]{@{}l@{}}Plays the chart title along with chart type.\\ Plays the X and Y axis labels.\\ Plays the chart description.\\ Plays the next or previous sentence.\\ Plays the currently selected sentence again.\\ Activate search mode. \\ Activate voice search mode. \\ To navigate through the data points.\\ To switch between lines in multi line chart.\\ To select multiple points for partial description.\\ To hear the list of selected points.\\ To reset the selected points.\\ To change the length of the description.\\ 
  Toggle audio speed
  \end{tabular} \\ \bottomrule
\end{tabular}}
\end{table}

\section{User Study}

\change{We conducted a user study with people with blindness to answer following questions. First, how the \textit{template-based} natural language summaries generated by our system fare in terms of user's performance compared to the \textit{human-written} summary condition (in which a human-written caption is provided)? Second, how the \textit{interactive summarization} feature 
affects user's performance to accomplish data exploration compared to when user manually examines each data point individually. Third, how do the users consider \seechart\ tool in terms of subjective measures (e.g., usefulness and ease of use)? Finally, what can we learn from participants' feedback to improve the future development of accessible visualization tools? 

In addition, we logged participants' key-strokes and the time they spent interacting with different features of the tool. We analyzed the log data to find the interaction patterns and most used features.}

\subsection{Participants}
We conducted the study with 14 participants (7 female, 7 male, age range 35-65, North-American) who had experience in using screen-reader tools (e.g., JAWS and NVDA). Six of them reported that they encounter charts occasionally (several times a month) and 8 of them said they encounter charts frequently (several times a week).

\subsection{Study Design} 
We designed the user study with  a $2 \times 2 \times 2$ 
within-subject design with following independent variables respectively: \textbf{Chart Type:} with 2 conditions (\textit{single bar}, \textit{multi line}), \textbf{Summary Type:} with 
\change{2 conditions (\textit{template-based}, \textit{human-written})}
and \textbf{Task type:} with 2 conditions (\textit{Task A}, \textit{Task B}). Here, \textit{Single Bar} refers to a bar chart with one categorical attribute and \textit{Multi Line} refers to line charts with multiple series. The dependent variables for the experiment were task completion time (measured in seconds) and percent of correct answers (\%). Each participant repeated each condition 2 times, thus, there were a total of 16 unique trials per participant. We counterbalanced the ordering of the conditions 
for each participant to reduce ordering effects. In total, in the study we collected a \change{total of 224 trials (14 participants $\times$ 2 chart types $\times$ 2 summary types $\times$ 2 task types $\times$ 2 trials).}


\subsection{Tasks}
We chose following two types of tasks based on the primitive analytic activity by Amar et al. \cite{amar2005low}. The first type of task involved finding extrema (\textit{Task A}). An example of \textit{Task A} is: ``Which industry maintained the highest share in GDP over the years?".

\begin{figure}[h]
\centering
  \includegraphics[width=.8\columnwidth]{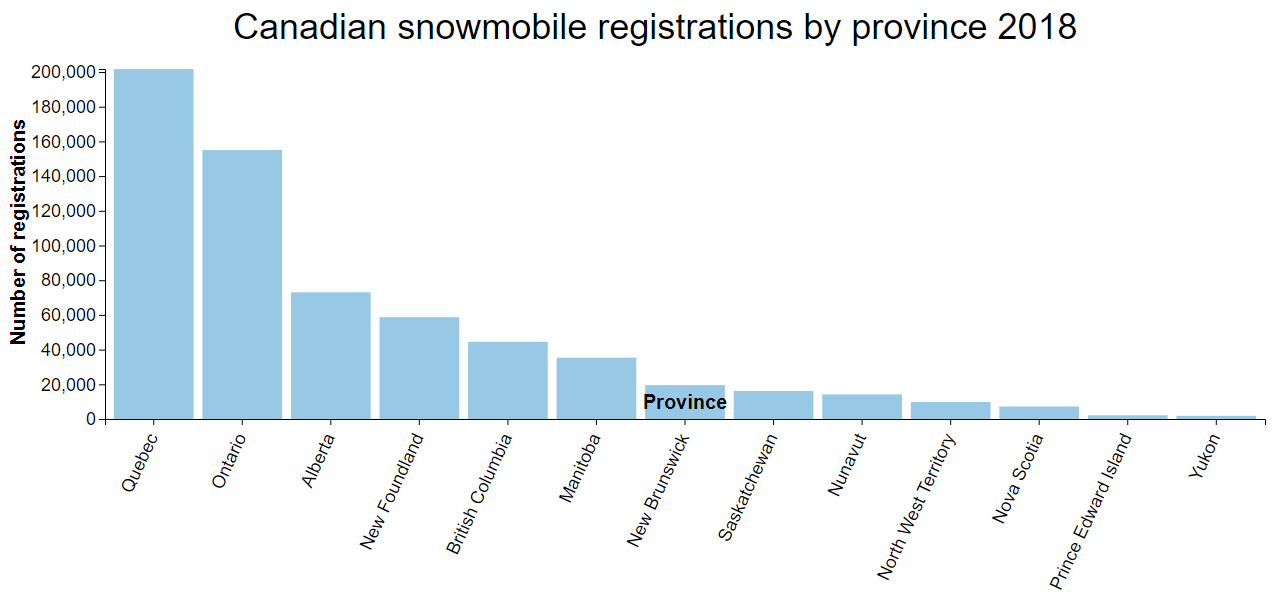}
   \vspace{-2mm}
  \caption{The bar chart example for Task B.}
  \label{fig:TASK_B}
\end{figure}

The second type of task is relatively complex as it combines two primitive tasks i.e., filtering following by finding extrema (\textit{Task B}). \change{Figure \ref{fig:TASK_B} shows the corresponding chart\cite{statista2} for the following question: ``Among the first five provinces shown in this chart, which province had the least number of snowmobile registrations?"}.
\change{8 out of the 16 trials presented \textit{Task A} while the other 8 trials presented \textit{Task B}.} 

\change{
For designing the tasks, we chose 16 different charts from Statista\cite{statista} in SVG format along with corresponding human-written summaries that are provided with the charts in Statista. Statista is an online platform that regularly publishes charts on a wide range of topics (e.g. transportation, economy, sports, opinion research, etc.). 
}


We chose charts with similar number of data items for each chart type to reduce the possible effects of confounding factors (data items for bar charts: \textit{M} $=$ 12.3, \textit{SD} $=$ 3.4, and for multi-line charts: \textit{M} $=$ 10.5, \textit{SD} $=$ 1.11).

\subsection{Procedure}


After filling up a pre-study questionnaire, participants went through a tutorial where they learned how to use the tool under different conditions and to complete the task. During the tutorial, we demonstrated the \seechart\ tool with sample charts that were not used during the main study. \change{In order to get comfortable with the tool, we allowed the participant to interact with it for 10-15 minutes.}

During the main study, each of the \change{16} trials presented a different chart along with a question. In the \textit{template-based} condition, the interface first presented the automatically generated moderate-length summary from our NLG method and then re-generated the summary whenever the user selected a range of data items. In the \textit{human-written} condition, participants had access to the human-written summary of the chart but \change{interactive summarization of selected data points was not available.}
In \change{both} conditions, participants could navigate through individual data points  using \texttt{arrow} keys. 
Participants explored the chart and then mentioned the answer out loud. 

After completing all \change{16} trials,
participants rated the \seechart\ tool on a 5-point Likert scale and expressed their opinions about the tool in a short interview. The experiment took about 90 minutes to complete and each participant received \$40.00 CAD for participating in the study. We developed an online experimental system to run the study which was accessed through the chrome browser. All study sessions were conducted remotely through video conference to conform to social distancing protocol due to COVID-19. 

\subsection{Results}

\subsubsection{Task Performance}

\begin{figure}[t!]
\centering
  \includegraphics[width=1\columnwidth]{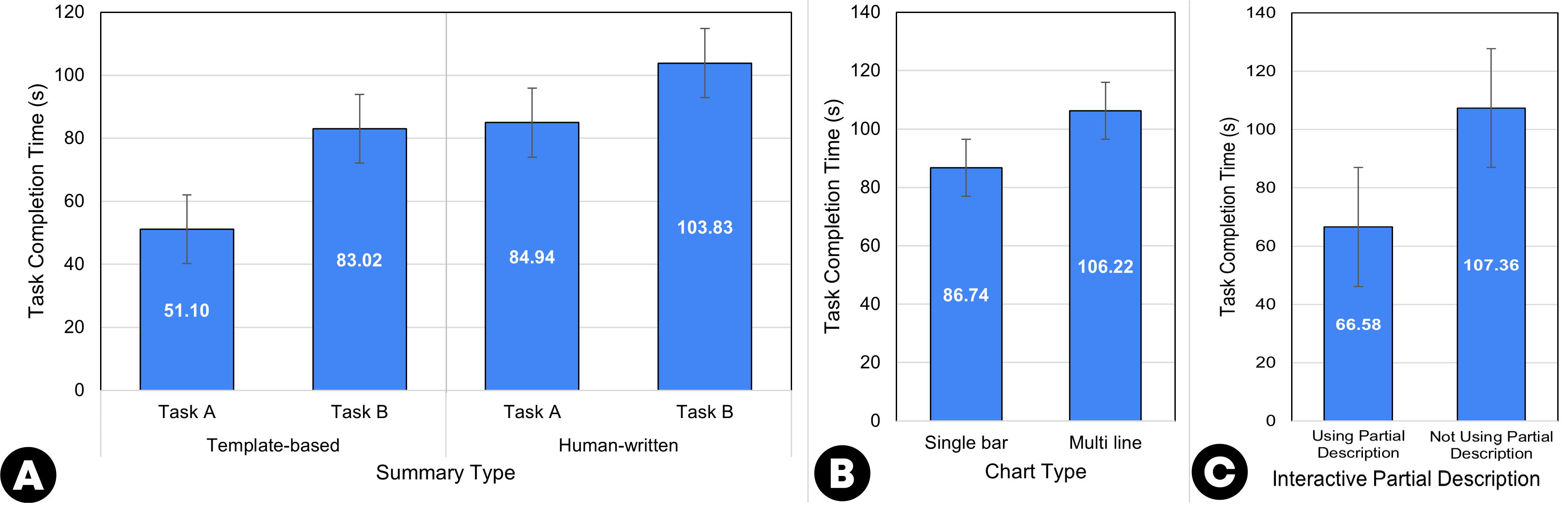}
  \caption{Mean task completion time for summary and task types (A), chart types (B), and application of interactive summarization.}
  \label{fig:_mean_time_for_chart_type}
\end{figure}

\textbf{Task Completion Time:} When we compare the summary types, we observe that participants were faster with \textit{template-based} summaries (\textit{M} $=$ 67.06, \textit{SD} $=$ 10.93) than \textit{human-written} (\textit{M} $=$ 94.39, \textit{SD} $=$ 13.83). \change{The results are shown in Figure \ref{fig:_mean_time_for_chart_type} (A). }
The analysis of variance (ANOVA) test revealed that the effect of \textit{summary type} on task completion time was \change{statistically significant ($F_{1, 13}$ $=$ 25.735, $p$ $<$ .0005)} 
while the effect of \textit{task type} on task completion time was not statistically significant \change{($F_{1, 13}$ $=$ 2.291, $p$ $>$ .05). }
One possible reason why participants were more efficient with template-based summaries is that they are more effective in answering questions by providing more consistent discourse structure and contents and focusing on statistical information rather than contextual and background information compared to human-written summaries. \change{Furthermore, participants completed the tasks faster (18.34\%) with single bar charts compared with multi line charts (Figure \ref{fig:_mean_time_for_chart_type} (B)). This suggests that users with blindness take more time with multi line charts possibly because  they are usually more complex than single bar charts.} 

\change{While accomplishing \textit{Task B}, users could apply the interactive summarization to get a partial summary of the chart to answer the questions in the \textit{template-based} condition which was not available in the \textit{human-written} condition and participants had to examine individual values of filtered data points to find the answers. Figure \ref{fig:_mean_time_for_chart_type}(C) shows the difference as participants were 37.98\% faster to complete \textit{Task B} when they could utilize the interactive summarization feature and the effect was statistically significant ($F_{1, 13}$ $=$ 15.320, $p$ $<$ .005).}

\begin{figure}[t!]
\centering
  \includegraphics[width=.8\columnwidth]{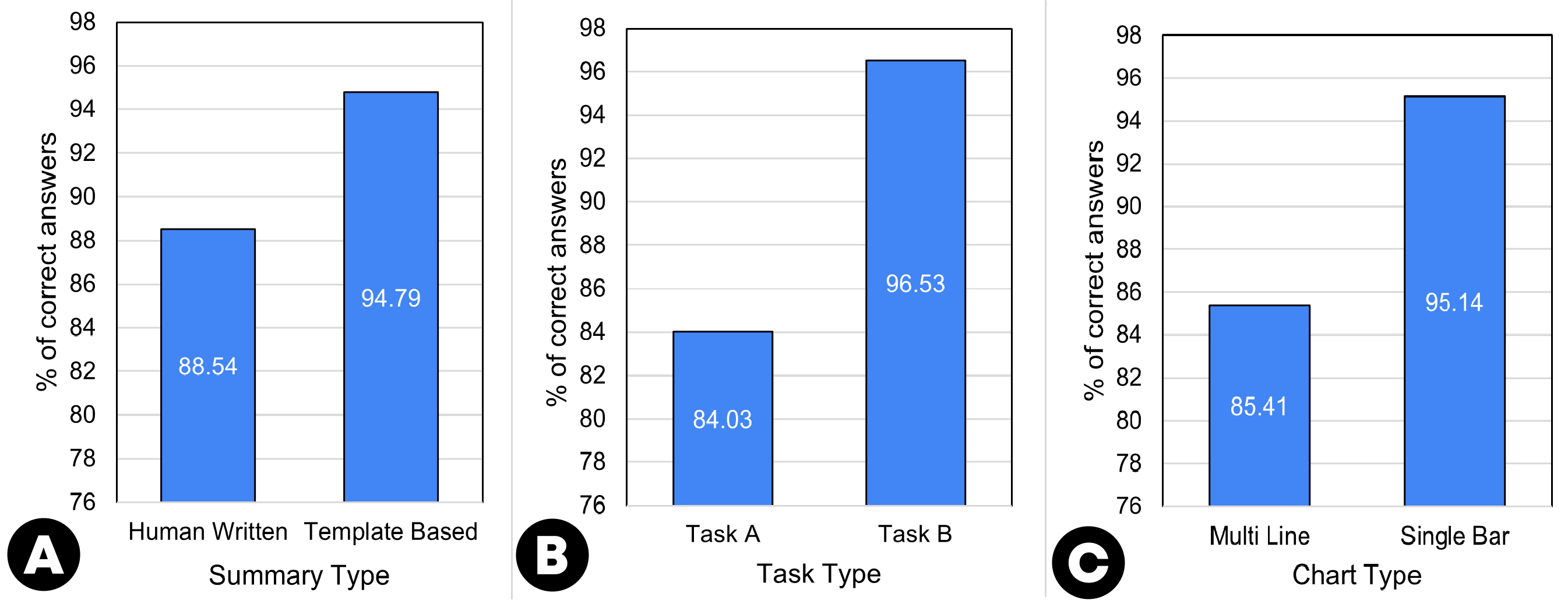}
   \caption{Percentage of correct answers based on summary type (A), task type (B), and chart type (C).}
   \label{fig:_accuracy}
 \end{figure}

\textbf{Accuracy:} Participants were also more accurate with \textit{template-based} summaries (94.79\%) than \textit{human-written} summaries (88.54\%) \change{as shown in Figure \ref{fig:_accuracy} (A). This indicates that human-written summaries are often generic and do not help to perform the common analytic tasks identified by Amar et al. \cite{amar2005low}. As a result, participants had to check the data points individually to find answers.} 
Interestingly, participants were less accurate with \textit{Task A} compared to \textit{Task B} (Figure \ref{fig:_accuracy} (B))
despite the fact that \textit{Task B} involves multiple operations (filtering followed by finding extrema) whereas \textit{Task A} involves finding extrema only. One possible explanation is that with \textit{Task A}, participants had to go through all the data points and keep track of their values to find extrema which can lead to a much more cognitive burden, than finding the extrema from a much smaller number of data points. Finally, more errors occurred with multi-line charts than with single-series bar charts (Figure \ref{fig:_accuracy} (C)), possibly because of the increasing complexity with more data series than single-series bar charts. \change{Eleven participants made at least one error when trials involved the \textit{human-written} condition, compared with 5 participants who made errors with the \textit{template-based} condition.}

\subsubsection{Post-study Feedback}
\begin{figure}[t!]
\centering
  \includegraphics[width=.8\columnwidth]{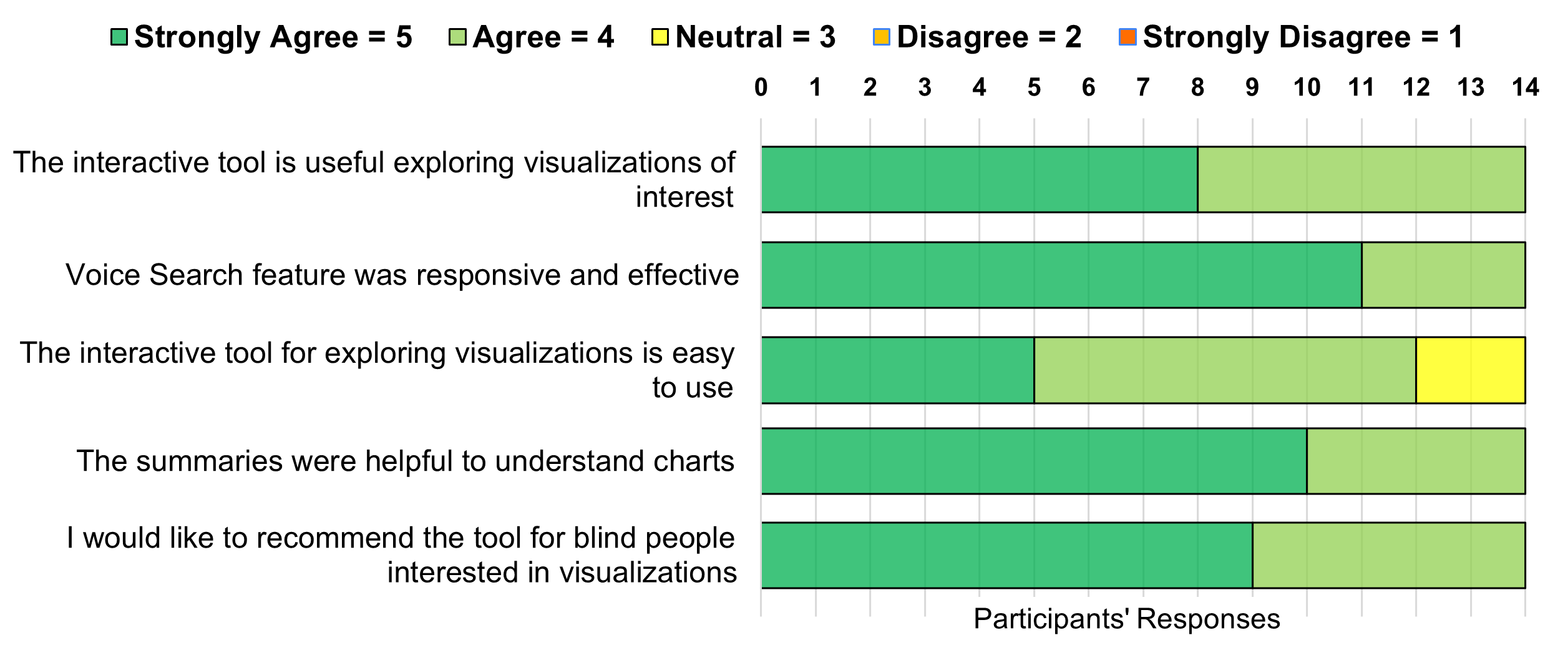}
   \caption{User study responses to post-study interview.}
   \label{fig:feedback}
 \end{figure}

At the end of the study, we asked participants for their feedback. 
Figure \ref{fig:feedback} shows the overall result of the Likert-scale questionnaire responses. As we can see all participants agreed or strongly agreed that the \seechart\ tool is useful for exploring visualizations and it can help them to understand charts. 
Also, they agreed or strongly agreed that they would recommend this tool to other people with visual impairments.

During interviews, all participants confirmed that previously they did not have any tool to help them interactively explore online charts through audio descriptions. Furthermore, there was no good way to get quick insights from charts based on selected points.  P09 said: ``\textit{This tool is so exciting. I never thought this would be possible to move across a chart and select data points to get such a detailed summary! Integration of the voice search just made life easier. Just like the other features, it was very responsive.}".
Most participants stated that they liked the interactive feature that generates summaries from the selected range of data points, as P03 said:
``\textit{This feature not only helps blind people to get insights from a selected region of a chart, but it will also be helpful to anyone interested in data analytics and visualization. 
}" 

Participants also liked the template-based summaries as they could get rich insights from charts that were not possible before. Eight participants liked the feature of accessing chart data with keyboard shortcuts. In general, they liked the audio feedback while navigating through the chart.  Three participants mentioned that they found the shortcut keys to be very conveniently designed and easy to remember.

\change{
Five of the participants mentioned that it would be extremely useful to have a search feature for getting statistically significant information (e.g., extrema, average, etc.) from the chart data. P06 said: ``\textit{I wish there was a quick search feature where I could ask the tool to find answers from the chart data rather than looking for it by myself}". The voice search feature (described in \ref{voice}) was later implemented and integrated into the \seechart\ tool.   
}

The interviews also revealed that while \seechart\ mostly
met their needs there were some concerns from participants. In particular, navigating through relatively complex charts such as the line chart with multiple series seemed difficult to at least four participants who reported that they were losing track of where they are as they were switching between different lines. Two of the participants suggested introducing a search feature so that they can quickly select specific data attributes or values without needing to navigate through all attributes and values. Further, one participant suggested to support  natural language question answering so that they can get their question answered through speech to perform simple analytical tasks such as sorting and filtering data.


\subsubsection{Interaction Patterns and feature usage}
\label{interaction}
\begin{figure}[t!]
\centering
  \includegraphics[width=.8\columnwidth]{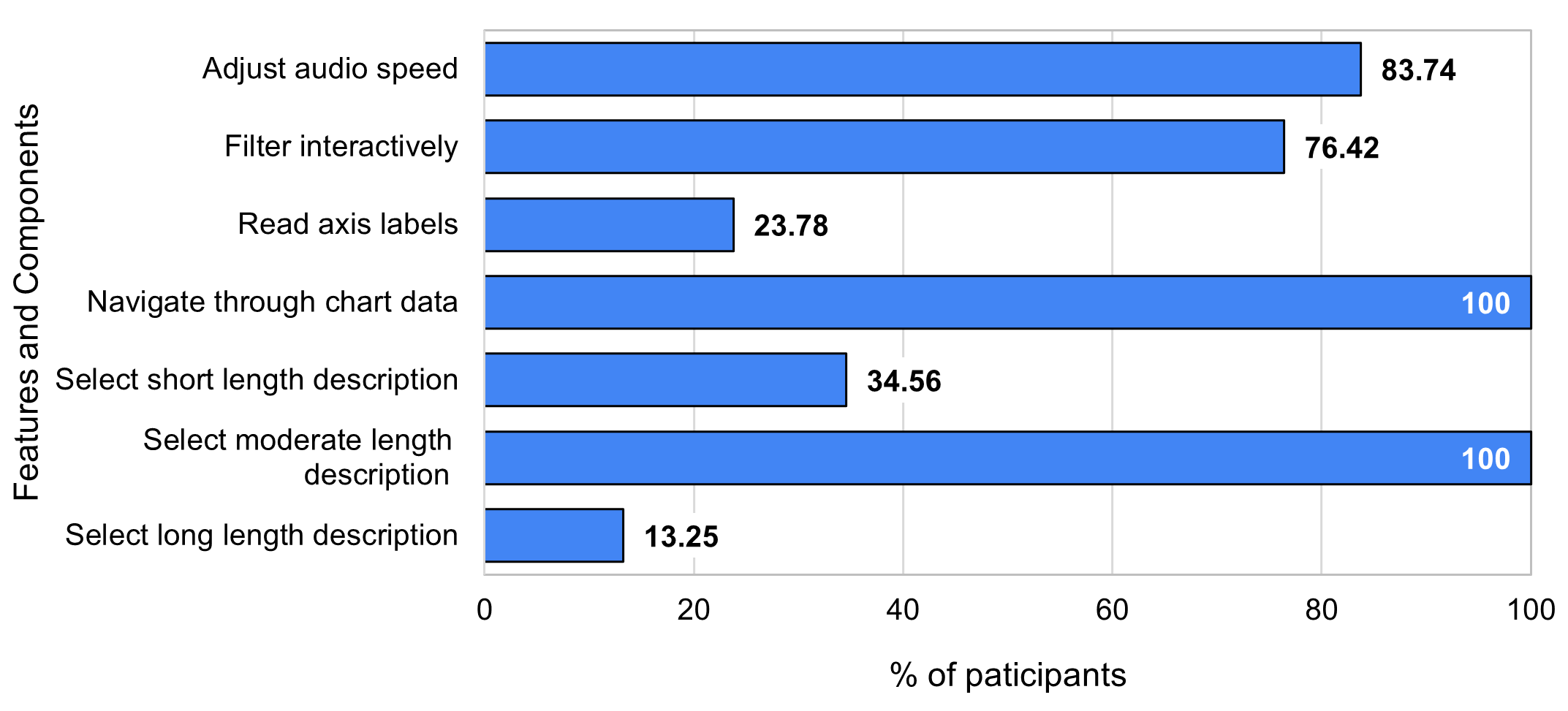}
   \caption{Usage of various features and components of \seechart\ tool.
    }
   \label{fig:interaction_feature}
 \end{figure}

\change{Besides questionnaires, we analyzed the log data to get insights into the interaction patterns of participants. As shown in Figure \ref{fig:interaction_sequence}, when a chart is presented to the participants, a majority (67\%) of them first listened to the chart description before going through the chart data and 33\% of the participants chose to hear the axis labels and chart data first before listening to the chart description. 

\begin{figure}[t!]
\centering
  \includegraphics[width=.8\columnwidth]{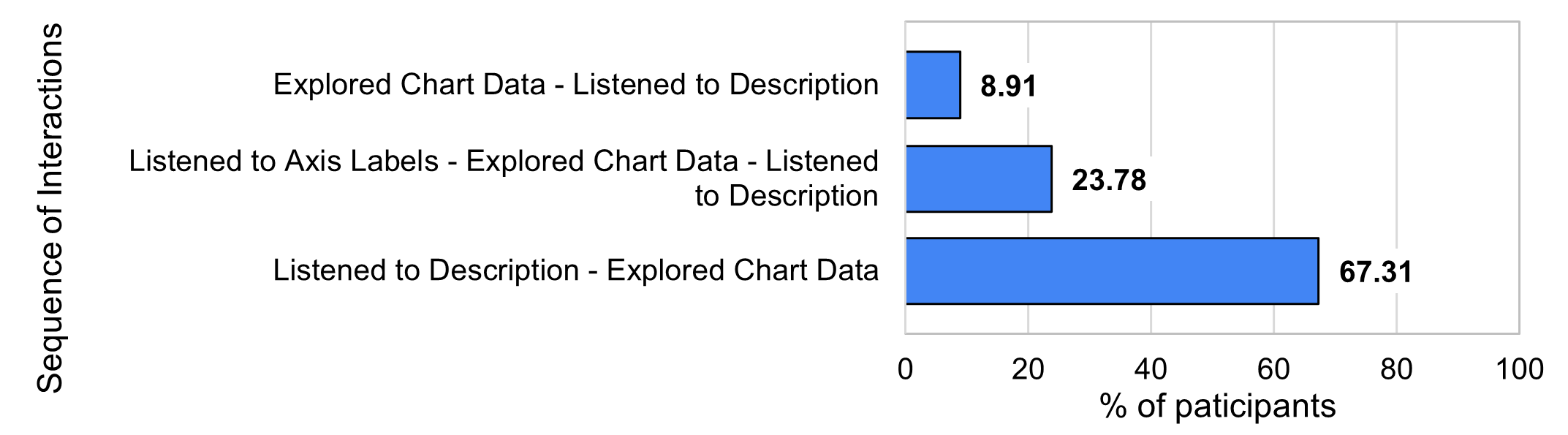}
   \caption{Sequence of interactions when a chart is presented to the participants.}
   \label{fig:interaction_sequence}
 \end{figure}
 
Figure \ref{fig:interaction_feature} shows the usage of various features and components of \seechart\ tool. As expected, all of the participants listened to the moderate length chart description and navigated through the chart data. 76\% participants utilized the interactive summarization feature to perform tasks. We found that very few participants (13\%) were interested to hear the long length descriptions and while 34\% of them listened to short length descriptions. A possible explanation is that the participants found the answer to the questions either by hearing the shorter length descriptions or checking the data values. Therefore, they did not require to listen to the long length descriptions. Axis labels were read by 23.78\% of participants who did not choose to listen to the descriptions first when a chart was presented. Participants who listened to the description first already knew what were the axis labels and they did not require to check the axis labels again. We also noticed that the default audio speed was too slow for 83.74\% of participants as they had to increase the tempo.

}  

\section{Limitations and Future Work}

The evaluation of  \seechart\ suggests that providing natural language summaries using template-based NLG methods is helpful in understanding charts on Web more efficiently and effectively. 
The study also revealed important lessons and helped us identify the limitations  that we would like to lift in future work. 

\textbf{Support complex charts through multiple modalities:} Currently, \seechart\ supports a variety of bar charts (e.g., single-series bar charts, stacked and grouped bar charts), line charts (with single or multiple series), and pie/donut charts. We chose these chart types in our initial prototype because they are frequently found on Web \cite{hoque-vis-search-2019} and can be deconstructed relatively easily. Moreover, existing chart summary corpora mainly cover these chart types \cite{accessibile-viz, chart-to-text}. In future, we aim to cover more complex charts such as mapchart, scatter-plot matrix and multiple coordinated views. However, with the increasing complexity of charts with larger number of data attributes and items, designing interactions can be more challenging as the participants struggle to keep track of multiple lines solely based on keyboard shortcuts. Future research may introduce additional input modalities such as touch and speech to perform brushing, filtering, and zoom in/out to support complex visualizations such as map charts, multiple coordinated views and improve scalability. Adopting and extending recent works on natural language interfaces for exploring and question-answering with data through speech can be starting point~\cite{hoque2017applying, srinivasan2018orko, kim2020answering}. In terms of output modalities, an interesting direction could be to combine data sonification along with audio summary based on interactions.

\textbf{Improve and extend chart deconstruction:}  \seechart\ currently supports SVG charts created using D3\cite{d3} and highcharts\cite{highchart} and can deconstruct basic charts like bar, line, pie charts, and scatter plots. In immediate future, we aim to support the extractions of SVG charts generated by other libraries such as Plotly\cite{plotly}, Chartblocks\cite{chartblocks}, and Graphiq~\cite{graphiq}.
Unlike chart images for which deconstructing data involves computer visualization challenges, recovering data and visual encodings can be relatively easier  from SVG-based charts by extracting features  such as position, sizes, and colors of SVG elements~\cite{battle2018beagle}. Similarly, Vega-Lite\cite{satyanarayan2017vega} charts provide declarative specifications which can be extracted~\cite{kim2020answering} for re-constructing the accessible version of a chart. While deconstructing image-based charts is still a challenging problem, more advanced deep learning methods have shown promise in extracting data and generating chart summaries from them~\cite{chart-to-text, obeid}. In future, such improved methods can be integrated with our \seechart\ tool to make these image-based charts more accessible.

\textbf{Explore advanced chart summarization methods:} During our design, in addition to the  template-based approach, we considered recently proposed deep learning methods based on the transformer architecture~\cite{chart-to-text, obeid}. while such advanced data-to-text NLG models usually generate fluent summaries and offer more variations in terms of reported insights, grammatical styles, and lexical choices than template-based summaries, they also suffer from  hallucinations (i.e., outputting irrelevant tokens) and factual errors. However, during our needfinding study participants preferred factually correct insights over rich variations in language, therefore we opted for utilizing template-based summaries. Our summarization method can also be improved in terms of improving the insight generation and more effective ways to rank the insights~\cite{wang2019datashot}.  However, generating more complex insights covering perceptual and cognitive phenomena is still an open problem~\cite{accessibile-viz}. Finally, our summaries do not cover contextual or domain-specific information that is not present within the given chart. However, this type of information has been found to be less important to the users than the insights covering statistics and data trends~\cite{accessibile-viz}.  Nevertheless, future work may exploit large-scale knowledge sources to inject domain knowledge into the summaries.


\final{
\textbf{Limitations of the user study:} Our initial evaluation of \seechart\ was conducted through user studies in a controlled environment with specific tasks and limited duration (90 minutes). As such, we limit the study to only two chart types (single bar, multi line) and collected limited amount of task performance and interaction log data. Similarly, we collected limited amount of subjective feedback with simple measures of usefulness and ease of use and did not use standard questionnaires such as SUS~\cite{sus} and NASA TLX~\cite{hart2006nasa}. Since user studies in controlled environment may lack realism~\cite{carpendale2008evaluating}, we plan to run a follow-up online longitudinal study by deploying the tool as a chrome extension among blind users which would allow us to measure adoption rate and gather further deeper insights.
}

\section{Ethical Considerations}
For the purpose of respecting the intellectual property of the charts that we presented during the evaluation, we only considered the publicly available charts complying with their terms and conditions. According to the publication rights of Statista\footnote{https://www.statista.com/getting-started/publishing-statista-content-terms-of-use-and-publication-rights}, users having campus license are given open access to the publicly available charts for academic purposes.


During the experiments, we anonymized the results and feedback to protect the privacy of our participants. In order to fairly compensate our participants, we considered more than the minimum hourly wage of Ontario, Canada which is 15.50\$. Our studies lasted 90-120 minutes and each participant received 40\$. 

Our template-based summaries can be misused to deceive users about the chart contents and their implications. The chart deconstruction process is greatly dependent on how the chart was designed and does not guarantee that the extracted data will be accurate at all times. As a result, \seechart\ may fail to reproduce the accessible version of the original chart precisely. 

To ensure the reproducibility of \seechart\ tool, the code repository has been made public. 

\section{Considerations for equity, diversity, and inclusions}
The primary goal of this work is to support people who are blind or visually impaired to explore visualizations on Web through a browser extension tool. Currently, 
people with visual disabilities are often unable to access charts which aggravates information inequalities~\cite{jung2021communicating}. Through this work, we aim to take an initial step towards making data visualization more accessible to people who are blind to address this inequality problem. Moreover, while the \seechart\ system is designed for people who are blind, in future it can also contribute towards making data visualization more inclusive and accessible among a diverse range of user populations by summarizing key insights from charts in lay language.


\section{Conclusions}
We have presented \seechart, an interactive web-based tool that enables accessible visualization through natural language description and keyboard-based interactions. The tool automatically deconstructs online charts and then generates an accessible version of the corresponding chart with automatically generated summaries and interactive features for exploring the charts. A user study with users with blindness confirms that such automatic summaries along with interactive features for exploration helped them read charts more efficiently and effectively. We hope that \seechart\ will contribute towards making information visualization and analytics more accessible through natural language as a modality and motivate other researchers to investigate this relatively unexplored area. \seechart\ is available at \url{https://github.com/vis-nlp/SeeChart}.

\bibliographystyle{ACM-Reference-Format}
\bibliography{sample-base}

\end{document}